\documentclass[aps,twocolumn,pra,superscriptaddress,nofootinbib]{revtex4-2}
\pdfoutput=1

\usepackage{amsmath}

\usepackage{amssymb}
\usepackage{amsthm}
\usepackage{mathrsfs}
\usepackage{bm, dsfont}
\usepackage[usenames,dvipsnames]{color}
\usepackage{colortbl}
\usepackage[table]{xcolor}
\usepackage{enumitem}
\usepackage{multirow}

\usepackage{natbib}
\usepackage[colorlinks=true,linkcolor=blue,citecolor=blue,urlcolor=blue]{hyperref}
\usepackage{cleveref}
\usepackage{hypcap}
\usepackage{verbatim, float}
\usepackage{psfrag}
\usepackage[normalem]{ulem}
\usepackage{physics}		% for braket, etc.
\usepackage{nicefrac}

\usepackage{listings}
\usepackage{graphicx} % [REVIEW] the 'demo' option was ON: it replaces EVERY figure by a black rectangle. Removed.

\usepackage{diagbox}	

\usepackage{relsize, scalerel}

\usepackage[caption=false]{subfig}

\newcommand{\be}{\begin{equation}}
\newcommand{\ee}{\end{equation}}

\newcommand{\id}{\mathds{1}}

\newcommand{\cE}{\mathcal {E}}

\definecolor{revgreen}{rgb}{0,0.45,0.15}

\newcommand{\new}[1]{{\color{blue}#1}}

\newtheorem{theorem}{Theorem}

\newtheorem{definition}{Definition}
\newtheorem{corollary}[theorem]{Corollary} % [REVIEW] was {Corollary}[theorem]; since no {theorem} is ever used, Corollary~\ref{cor: WP} printed as "Corollary 0.1".

\newtheorem{proposition}[theorem]{Proposition}

\newtheorem*{remark}{Remark}

\begin{document}

\title{The bottleneck dimension of quantum operations} % [REVIEW] dropped the trailing period.

\author{Pavel Sekatski}
\affiliation{Department of Applied Physics, University of Geneva, Switzerland}
\begin{abstract}
Programmable quantum devices nominally act on a Hilbert space whose dimension grows exponentially with the number of constituents, but the presence of noise makes it unlikely that they remain coherent across all of this immense Hilbert space. Then, what is the effective coherent quantum dimension that should be associated with such imperfect devices?
To answer this question we here introduce an operational basis-independent framework which imposes a {\it dimension bottleneck} on the programmable transformations. Concretely we ask how strongly the quantum information they process can be compressed. Formalizing this idea we identify three inequivalent notions, termed $d$-compressibility, $d$-simulability and $d$-embeddability, which differ in the causal structure used to impose the bottleneck and form a strict hierarchy. 
The framework unifies several existing notions: joint measurability and simulability of quantum measurements, and the absolute dimensionality of state ensembles, are recovered as special cases. We illustrate the hierarchy with noisy qubit measurements in complementary bases, and we determine the white-noise thresholds at which the set of all noisy unitary channels in dimension $n$, a noisy universal quantum processor,  becomes $d$-compressible, $d$-simulable and $d$-embeddable. The thresholds confirm the expectation — maintaining coherence across the full Hilbert space becomes increasingly demanding as the nominal dimension $n$ increases.
\end{abstract}
\maketitle

\section{Introduction}

Past decades have witnessed tremendous progress in our capability to coherently manipulate larger and larger quantum systems. Today everyone has the opportunity to remotely control a programmable quantum device acting on several tens of qubits, that is a quantum system of dimension counted in trillions. Despite this progress, the operations implemented on such platforms remain inherently noisy, and it is unlikely that coherence is maintained across all of this immense Hilbert space. This raises a fundamental question: how can one rigorously characterize when a programmable quantum transformation acts coherently on its entire nominal Hilbert space? More generally, what is the minimal intrinsic quantum dimension on which the transformation acts coherently?

Motivated by these questions, in this work we introduce an operational framework to quantify the intrinsic dimension of a set of quantum operations, through the perspective of their maximal compression. Our approach is grounded in the idea of a dimensional bottleneck --
the smallest Hilbert space dimension to which quantum information can be compressed while still allowing the target operations to be implemented faithfully. By formalizing this idea, we provide a principled way to assess the effective bottleneck dimension of quantum processes, beyond the nominal size of the associated Hilbert space.

More concretely, in Sec.~\ref{sec:definition} for a general programmable transformation described by a set of quantum instruments we define three nonequivalent notions, termed $d$-compressibility, $d$-simulability and $d$-embeddability, determined by the precise causal structure used to impose the dimension bottleneck. We show that for a set of instruments $d$-embeddability implies $d$-simulability, which implies $d$-compressibility; with a strict separation between the three. We find that the first, least restrictive, notion of $d$-compressibility is directly related to the Schmidt number~\cite{huang2006schmidt} of the instruments, and is the only one which does not depend on {\it the relation} between different instruments in the set.

In Sec.~\ref{sec: special cases} we discuss some special cases. First, we show that when restricting to the sets of instruments without a quantum output (measurements) and instruments without a quantum input (states), $d$-simulability and $d$-embeddability recover two definitions that have been proposed previously in Refs.~\cite{ioannou2022simulability} and~\cite{bernal2024absolute}; thus unifying the two under a more general framework. Second, we focus on the case of trivial $(d=1)$ bottleneck dimension, arguing that a set is $1$-compressible if and only if it is composed of entanglement-breaking maps. In turn, $1$-simulability can be understood as the maps being jointly-entanglement-breaking, while $1$-embeddability is uninteresting. 

Finally, in Sec.~\ref{sec: examples} we discuss a series of examples: in particular, two-instrument sets describing noisy qubit measurements in the $X$ and $Z$ bases, as well as the set of all noisy unitary operations in dimension $n$, which can be thought of as a (noisy) universal quantum processor.

\section{Definitions}

\label{sec:definition}
Consider a programmable quantum device (or processor) that transforms a quantum system of dimension $n$ to a quantum system of dimension $m$. The programmable aspect can be captured by introducing a classical variable $x$, that the device takes as input. There is no good reason to forbid our processor to perform intermediate measurements on the quantum system, hence we also allow it to have a classical output $a$. In quantum physics such a device is described by a set of quantum instruments $\{\mathcal{I}_{a|x}\}$: a collection of completely positive (CP) linear maps 
\be
\mathcal{I}_{a|x} : L(\mathds{C}^n) \to L(\mathds{C}^m)
\ee
indexed by the input $x$ and the output $a$, such that  $\sum_{a} \mathcal{I}_{a|x}$ is also trace preserving (TP) for all $x$.

\begin{figure}[b]
    \centering
    \includegraphics[height=1.3cm]{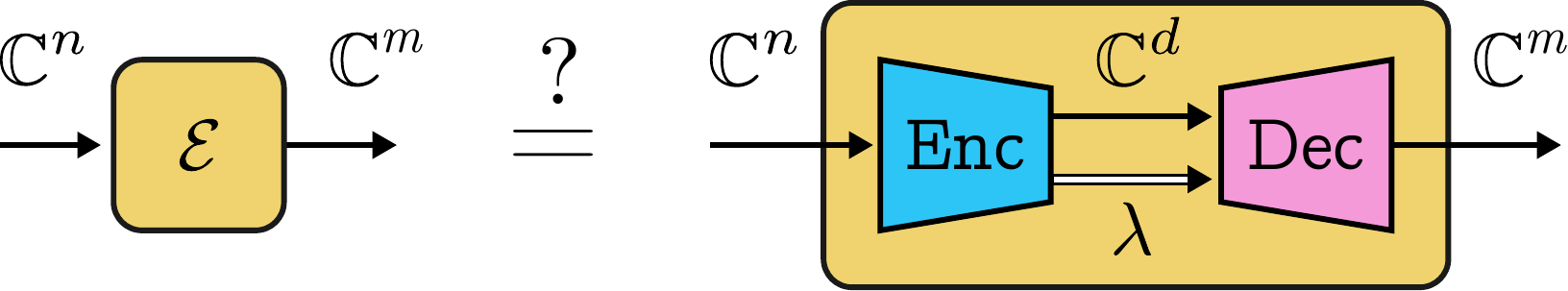}
    \caption{A quantum  channel $\cE$, mapping a $n$-dimensional quantum system to a $m$-dimensional one, is compressible to dimension $d$ if it can be decomposed as depicted in the figure, where $\lambda$ is a classical variable. This is the case iff the channel has Schmidt number ${\rm SN}(\cE)\leq d$. Here and below we follow the standard notation for circuit diagrams: full lines represent quantum systems whose dimension is specified above, while double lines represent classical systems.}
    \label{fig: generalchan} 
\end{figure}

Although the instruments nominally operate on a quantum system of dimension $n$, it is easy to think of examples where they describe classical processes. More generally one should not expect noisy operations to have a genuinely coherent behavior across the whole Hilbert space. To assess the minimal dimension where the set of instruments is irreducibly quantum, we will follow the intuition of the {\it dimension bottleneck}.
\vspace{2 mm}

\noindent
{{\bf Intuition.} \it The operations $\{\mathcal{I}_{a|x}\}$ are not genuinely $(d+1)$-dimensional quantum if they admit a realization where all quantum information is, at some point, compressed to a system of dimension $d$.}
\vspace{2 mm}

\noindent
We will now make this intuition mathematically precise, starting with a simple example of a single quantum channel (CPTP map).

\subsection{Warm-up: dimension bottleneck for a single channel}

% \begin{figure*}[t!]
%     \centering
%     \includegraphics[height=2.16cm]{IntrumentsEqv2.pdf}
%     \caption{A quantum instrument $\{\mathcal{I}_a\}$ acting on a $n$-dimensional quantum system is compressible to dimension $d$ if admit one of the three (equivalent) decompositions where $\lambda$ is a classical variable. Note that in $(iii)$ the decoder recieves the classical variable $\lambda$ directly from the encoder, inaffected by the latent instument}
%     \label{fig: generalinst} 
% \end{figure*}

To start gently consider  a single quantum channel $\cE$ transforming the quantum system of dimension $n$ to the one of dimension $m$. To pass the quantum information through a dimension bottleneck it must first encode the input in a system of dimension $d$ and then decode it back to the $m$-dimensional output. Nevertheless, by doing so it should be able to transmit classical information for free --- it would be absurd to associate a high-capacity {\it classical} channel with a nontrivial quantum dimension.  This leaves us with a single meaningful definition.
\vspace{2 mm}

\noindent
{\it
 A quantum channel $\cE : L(\mathds{C}^n)\to L(\mathds{C}^m) $ is called { $d$-compressible}    (Fig.~\ref{fig: generalchan}) if it admits the decomposition
\begin{align}\label{eq: 1 channel} \cE = \sum_\lambda {\tt Dec}_{|\lambda} \circ {\tt Enc}_{\lambda},
\end{align}
where $\{{\tt Enc}_{\lambda}\}$ is an instrument that outputs a $d$-dimensional quantum system and a classical variable $\lambda$, while $\{{\tt Dec}_{|\lambda}\}$ is a set of channels controlled by $\lambda$.
}
\vspace{2 mm}

It is easy to see, that $d$-compressibility of a channel is directly related to its {\it Schmidt number}.
\begin{definition} \textit{\textbf{The Schmidt number}}~\cite{huang2006schmidt}, ${\rm SN}(\cE)$, of a CP map $\cE$ is the minimal integer $d$ for which the map admits a Kraus decomposition $\cE [\cdot]= \sum_\lambda K_\lambda \cdot K_\lambda^\dag$ with ${\rm rank}(K_\lambda)\leq d$; it is equal to the Schmidt number~\cite{terhal2000schmidt} of the associated Choi operator $J_\cE:=({\rm id}\otimes \cE)\left[\sum_{i,j=1}^n \ketbra{ii}{jj}\right]$.  
\label{def: SN}
\end{definition}

% \noindent
% {\it
%  {\bf The Schmidt number} ${\rm SN}(\cE)$ of a CP map $\cE : L(\mathds{C}^n)\to L(\mathds{C}^m)$ is the minimal integer $d$ for which the channel admits a Kraus decomposition $\cE [\cdot]= \sum_\lambda K_\lambda \cdot K_\lambda^\dag$ with ${\rm rank}(K_\lambda)\leq d$. It is equal to the Schmidt number [...] of the assotiated Choi operator $J_\cE:=({\rm id}\otimes \cE)\left[\sum_{i,j=1}^n \ketbra{ii}{jj}\right]$.
% }
% \vspace{2 mm}

\noindent To see the tight relation, first take the Kraus representation $\cE [\cdot]= \sum_\lambda K_\lambda \cdot K_\lambda^\dag$, imposed by the Schmidt number bound, 
and apply the singular value  decomposition to each Kraus operator of rank $d$. We get $K_\lambda = V_\lambda \Pi^{(d)} M_\lambda U_\lambda$, where $\Pi^{(d)}$ is a projector on a fixed $d$-dimensional subspace. Hence, the Kraus operators $D_\lambda :=V_\lambda \Pi^{(d)}$ and  $E_\lambda :=\Pi^{(d)} M_\lambda U_\lambda$ readily define the operations ${\tt Dec}_{|\lambda}$ and ${\tt Enc}_\lambda$ respectively (note that $\sum_\lambda E_\lambda^\dag E_\lambda = \sum_\lambda K_\lambda^\dag K_\lambda = \id$). Conversely, Eq.~\eqref{eq: 1 channel} endows the channel with a Kraus decomposition $\cE [\cdot]= \sum_{\lambda,i,j} K_{\lambda,i,j} \cdot K_{\lambda,i,j}^\dag$ with $K_{\lambda,i,j}= D_{\lambda,i}E_{\lambda,j}$ satisfying ${\rm rank}(K_{\lambda,i,j})\leq d$.  Hence, with our definition we have just recovered the standard notion of the Schmidt number of a channel.
\vspace{2 mm}

\noindent
\begin{center}
{\it A channel $\cE$ is $d$-compressible iff it has ${\rm SN}(\cE)\leq d$}.
\end{center}

\subsection{Dimension bottleneck for a set of instruments}

\begin{figure}[t!]
    \centering
    \includegraphics[height=2.1 cm]{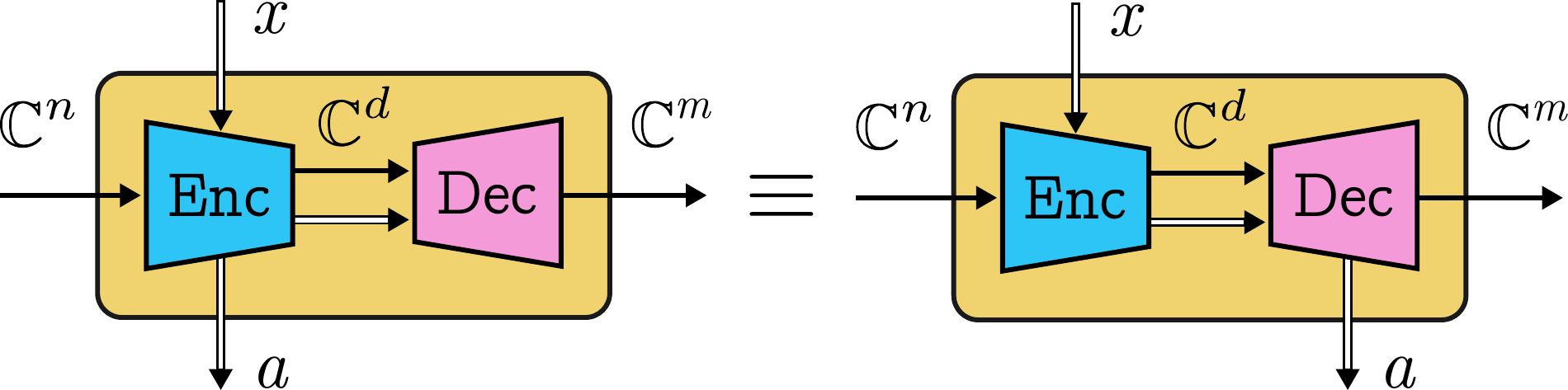}
    \caption{ A \textit{\textbf{$\bm d$-compressible}} set of instruments $\{\mathcal{I}_{a|x}\}$ admits one of the equivalent decompositions depicted in the figure.
    % Different decompositions of a set of quantum instruments $\{\mathcal{I}_{a|x}\}$, formalizing the intuition of the {\it dimension bottleneck $d$}. Note that for the {\it d-embeddable} decomposition the decoder $\{{\tt Dec}_{|\lambda}\}$ receives the classical variable $\lambda$ directly from the encoder $\{{\tt Enc}_{\lambda}\}$, bypassing the latent instrument $\{{\tt Lat}_{a|\lambda,x}\}$. \ps{Without this restriction the $d$-embeddable and $d$-simulable decompositions would coincide.}
    }
    \label{fig: dcompress} 
\end{figure}

Let us now come back to our general set of quantum instruments $\{\mathcal{I}_{a|x}\}$, and use the example of the channel as a guideline. This case is richer in many aspects. A particularly important one is that the set contains several transformations, labeled by the input $x$,  and {\it the relation} between them can play a key role. 

Practically speaking, to adjust our channel definition of compressibility to the instrument set, we must figure out where exactly in the diagram of Fig.~\ref{fig: generalchan} the classical variables $x$ and $a$ are to be respectively input and output; with the obvious condition that the input must causally precede the output.
There are essentially three possibilities. {\it (i)} The classical variables can be manipulated by the encoder.  {\it (ii)} The classical variables can be manipulated by the decoder. {\it (iii)} The classical variables can be manipulated by a latent instrument acting on the compressed $d$-dimensional quantum system. We now discuss these possibilities one after the other\\
\vspace{2 mm}

\begin{definition}
 A set of quantum  instruments $\{\mathcal{I}_{a|x} \}$, with $\mathcal{I}_{a|x} : L(\mathds{C}^n)\to L(\mathds{C}^m) $, is called \textit{\textbf{$\bm d$-compressible}}    (Fig.~\ref{fig: dcompress}) if it admits either of the equivalent decompositions
\begin{align}
\label{eq: set compressible 1} 
\mathcal{I}_{a|x} &=
\sum_\lambda {\tt Dec}_{|\lambda} \circ {\tt Enc}_{a,\lambda|x}\\
\label{eq: set compressible 2} 
\mathcal{I}_{a|x} &=
\sum_\lambda {\tt Dec}_{a|\lambda} \circ {\tt Enc}_{\lambda|x},
\end{align}
where the encoder (resp. decoder) outputs (resp. inputs) a quantum system of dimension $d$, in addition to the classical variables. 
%\oldm{A set is $d$-compressible iff ${\rm SN}(\mathcal{I}_{a|x})\leq d$ for all CP maps.}
\end{definition}

\begin{remark}
A set of instruments is $d$-compressible if and only if ${\rm SN}(\mathcal{I}_{a|x})\leq d$ for all $a$ and $x$.
\end{remark}

\noindent 
We first prove that the decompositions in Eqs.~\eqref{eq: set compressible 1} and \eqref{eq: set compressible 2} are equivalent. The direction \eqref{eq: set compressible 1} $\implies$ \eqref{eq: set compressible 2} is immediate, since an outcome $a$ produced at the encoder can always be passed to the decoder via the classical variable, and revealed there.
To show the other direction, consider a general Kraus decomposition of the decoder in Eq.~\eqref{eq: set compressible 2}
\be
{\tt Dec}_{a|\lambda}[\cdot] = \sum_i  D_{a,i|\lambda}\cdot  D_{a,i|\lambda}^\dag.
\ee
With the polar decomposition we can write each of the Kraus operators as $D_{a,i|\lambda}=U_{a,i,\lambda} D_{a,i|\lambda}^{(d)},$
where $U_{a,i,\lambda}$ are semi-unitary $m\times d$ matrices defining the embedding from dimension $d$ to dimension $m$, and $D_{a,i|\lambda}^{(d)}$ are positive semidefinite $d\times d$ matrices defining an instrument on the $d$-dimensional system  ($\sum_{i,a}(D_{a,i|\lambda}^{(d)})^2= \id$). We can thus split these two steps as two physically separate operations, and move the $d$-dimensional instrument, given by the Kraus operators $\{D_{a,i|\lambda}^{(d)}\}$, to the encoder. Defining the new classical variable $\lambda'=(\lambda,a,i)$ sent from the encoder to the decoder we obtain the decomposition in the Eq.~\eqref{eq: set compressible 1}, and prove \eqref{eq: set compressible 2} $\implies$ \eqref{eq: set compressible 1}.

Remark, that the definition of $d$-compressibility is a straightforward {\it element-wise} generalization of the one-instrument notion to the set, and does not depend on the relation between different instruments. Since the label $x$ is given from the start and can be fed forward, the compression of each instrument from the set can be done independently from the others. Hence, the full set is $d$-compressible if and only if each instrument in the set is. But for a single instrument $d$-compressibility is equivalent to all CP maps composing it having a Schmidt number bounded by $d$. The proof is a straightforward adaptation of the CPTP map case, given in the previous section.\\

Next, we consider decompositions, where the input $x$ is only provided at the decoder. As we shall see, such decompositions can be equivalently defined by requiring that the variables $x$ and $a$ be provided by a latent instrument, inserted between the encoder and the decoder and acting on the intermediate $d$-dimensional quantum system.
This gives rise to the following definition.

\begin{figure}[t!]
    \centering
    \includegraphics[width=\columnwidth]{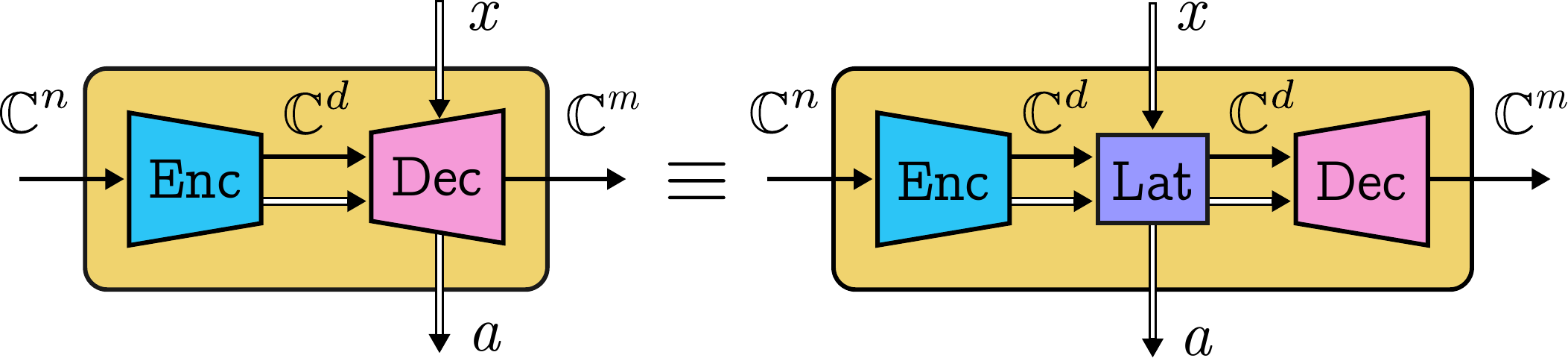}
    \caption{ A \textit{\textbf{$\bm d$-simulable}} set of instruments $\{\mathcal{I}_{a|x}\}$ admits one of the equivalent decompositions depicted in the figure.
    % Different decompositions of a set of quantum instruments $\{\mathcal{I}_{a|x}\}$, formalizing the intuition of the {\it dimension bottleneck $d$}. Note that for the {\it d-embeddable} decomposition the decoder $\{{\tt Dec}_{|\lambda}\}$ receives the classical variable $\lambda$ directly from the encoder $\{{\tt Enc}_{\lambda}\}$, bypassing the latent instrument $\{{\tt Lat}_{a|\lambda,x}\}$. \ps{Without this restriction the $d$-embeddable and $d$-simulable decompositions would coincide.}
    }
    \label{fig: dsimul} 
\end{figure}

\vspace{2 mm}
\begin{definition}
 A set of quantum instruments $\{\mathcal{I}_{a|x}\}$ with $\mathcal{I}_{a|x} : L(\mathds{C}^n)\to L(\mathds{C}^m) $ is called \textit{\textbf{$\bm d$-simulable}}    (Fig.~\ref{fig: dsimul}) if it admits either of the equivalent decompositions
\begin{align}
\label{eq: set simmulable 1} 
\mathcal{I}_{a|x} &= 
\sum_\lambda {\tt Dec}_{a|\lambda,x} \circ {\tt Enc}_{\lambda}\\
\label{eq: set simmulable 2} 
\mathcal{I}_{a|x} &=  \sum_{\lambda,\lambda'} {\tt Dec}_{|\lambda'} \circ {\tt Lat}_{a,\lambda'|\lambda,x} \circ {\tt Enc}_{\lambda}
\end{align}
where the encoder and the latent instrument output a quantum system of dimension $d$, while the decoder and the latent instrument take as input a quantum system of dimension $d$, in addition to the classical variables.
\end{definition}

The proof of the equivalence of the two decompositions is very similar to the above.
The implication \eqref{eq: set simmulable 2}$\implies$\eqref{eq: set simmulable 1} is immediate, since we can view ${\tt Dec}_{|\lambda'} \circ {\tt Lat}_{a,\lambda'|\lambda,x}$ as a single decoding operation. For the converse direction, we again use the polar decomposition of the Kraus operators representing the decoder ${\tt Dec}_{a|\lambda,x}$ in Eq.~\eqref{eq: set simmulable 1}
\be
{\tt Dec}_{a|\lambda,x }[\cdot] = \sum_i  D_{a,i|\lambda,x}\cdot  D_{a,i|\lambda,x}^\dag,
\ee
with $D_{a,i|\lambda,x}=U_{a,i,\lambda,x} D_{a,i|\lambda,x}^{(d)}$. Here $\{D_{a,i|\lambda,x}^{(d)}\}$ define a set of instruments on the $d$-dimensional system, which we identify with the latent instruments $\{{\tt Lat}_{a,\lambda'|\lambda,x}\}$ in Eq.~\eqref{eq: set simmulable 2} for $\lambda':=(i,\lambda, a,x)$, showing \eqref{eq: set simmulable 1}$\implies$\eqref{eq: set simmulable 2}. 

When comparing the definition of $d$-simulability (Eq.~\ref{eq: set simmulable 1}, Fig.~\ref{fig: dsimul}) with that of $d$-compressibility (Eq.~\ref{eq: set compressible 2}, Fig.~\ref{fig: dcompress}), it is intuitively clear that providing the input $x$ later must be more constraining for a decomposition. Indeed, we see that any $d$-simulable set $\{\mathcal{I}_{a|x}\}$ is $d$-compressible, since in the latter case the input $x$, provided at the encoder, could always be transmitted to the decoder via the classical variable. To see that the converse is not true, think of a set of quantum measurements, i.e. instruments without a quantum output or, in other words, with a  trivial output dimension $m=1$. Any set of measurements is trivially $1$-compressible; however, it is $1$-simulable if and only if it is {\it jointly measurable}~\cite{busch2016quantum,heinosaari2016invitation}, which is a nontrivial property. There is thus a strict inclusion of the class of $d$-simulable sets of instruments in the class of $d$-compressible ones.\\
%, as illustrated in Fig.~\ref{fig: venn} \\
%%%%%%%%%%%%%%%%%%%%%%%%%%%%%%%%%%%%%%%%%%%%%%%%%%

% \vspace{2 mm}
% \begin{center}
% {\it  d-compressible $\quad \supsetneq \quad$ d-simulable.}
% \end{center}
% \vspace{2 mm}

Finally, the decomposition with the latent instrument can be modified by constraining the classical information exchanged between the latent instrument and the decoder. This leads us to introduce the following definition.

\begin{figure}[t!]
    \centering
    \includegraphics[height=2 cm]{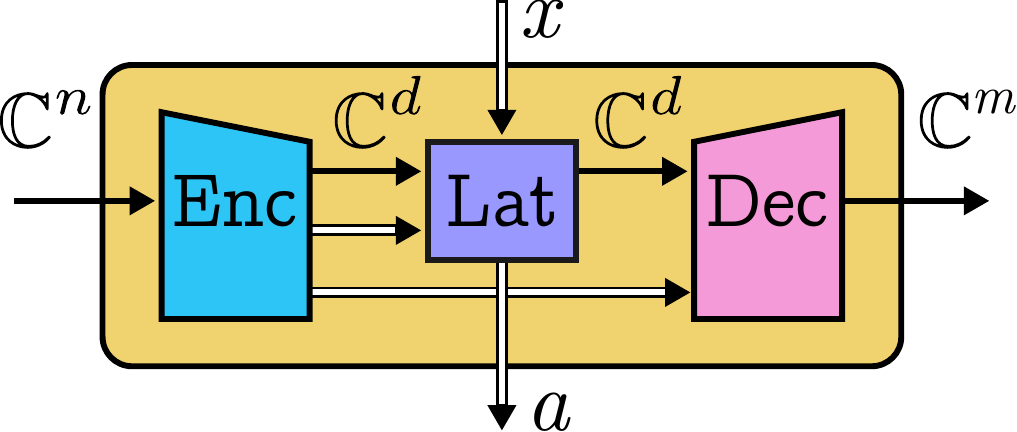}
    \caption{ A \textit{\textbf{$\bm d$-embeddable}} set of instruments $\{\mathcal{I}_{a|x}\}$ admits  the decomposition depicted in the figure.
    % Different decompositions of a set of quantum instruments $\{\mathcal{I}_{a|x}\}$, formalizing the intuition of the {\it dimension bottleneck $d$}. Note that for the {\it d-embeddable} decomposition the decoder $\{{\tt Dec}_{|\lambda}\}$ receives the classical variable $\lambda$ directly from the encoder $\{{\tt Enc}_{\lambda}\}$, bypassing the latent instrument $\{{\tt Lat}_{a|\lambda,x}\}$. \ps{Without this restriction the $d$-embeddable and $d$-simulable decompositions would coincide.}
    }
    \label{fig: demb} 
\end{figure}

\begin{definition} A set of quantum instruments $\{\mathcal{I}_{a|x}\}$ with $\mathcal{I}_{a|x} : L(\mathds{C}^n)\to L(\mathds{C}^m) $
is called \textit{\textbf{$\bm d$-embeddable}} (Fig.~\ref{fig: demb})  if it admits  the decomposition
\begin{align}
\label{eq: set embedable}
\mathcal{I}_{a|x} &=  \sum_\lambda {\tt Dec}_{|\lambda} \circ {\tt Lat}_{a|\lambda,x} \circ {\tt Enc}_{\lambda},
\end{align}
where the encoder and  the latent instrument output a quantum system of dimension $d$, while the decoder and the latent instrument take as input a quantum system of dimension $d$, in addition to the classical variables.
\end{definition}

This definition is obtained from $d$-simulability by forbidding the classical communication\footnote{An attentive reader has noticed that restricting the classical information exchange differently, gives two more possible decompositions of the instruments set. These variants are briefly discussed in Appendix~\ref{app: d-imbed}.} from the latent instrument to the decoder. It is thus clear that $d$-embeddability implies $d$-simulability. To see that the converse is not true consider the sets of quantum states, i.e.\ instruments with trivial input dimension $n=1$. Since there is no quantum input any set of states is trivially $1$-simulable. In turn, a set of two (nonidentical) states is not $1$-embeddable, since for $d=1$ there would be no causal relation between the input label $x$ and the output quantum state. 

\begin{figure}[t!]
    \centering
    \includegraphics[width=0.85\columnwidth]{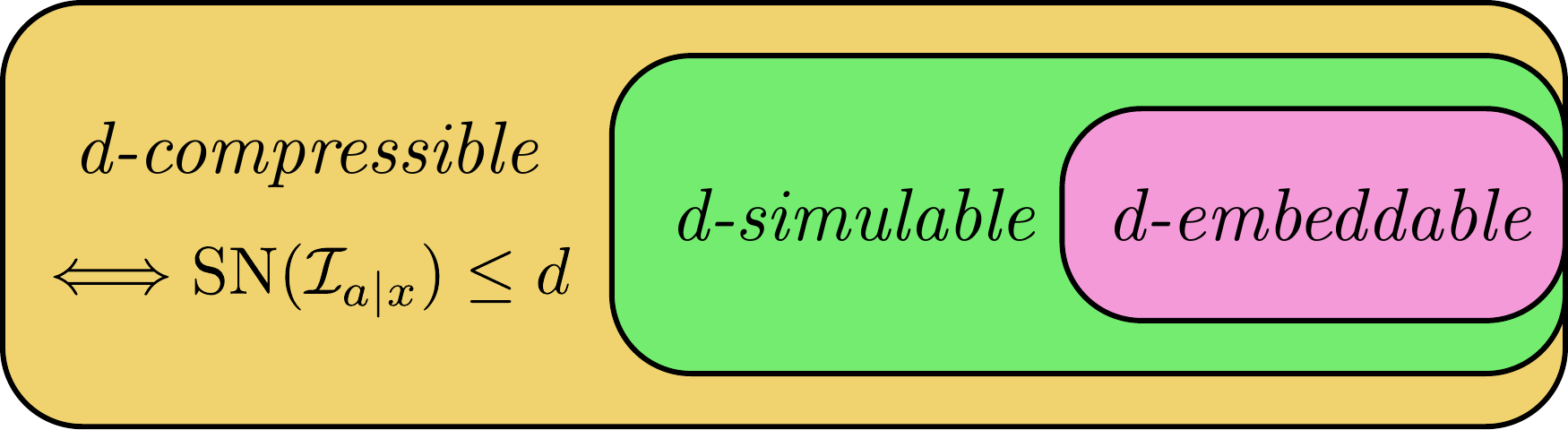}
    \caption{
    Hierarchy of the dimension-bottleneck properties. The sets of instrument families $\{\mathcal{I}_{a|x}\}$ that are 
$d$-embeddable, 
$d$-simulable and 
$d$-compressible are strictly nested.
    }
    \label{fig: venn} 
\end{figure}

The relation between the three dimension-bottleneck notions is summarized  in Fig.~\ref{fig: venn}. We recall that, in contrast to $d$-compressibility, the other two notions depend on the relation between the different instruments in the set. In fact, for a set consisting of a single instrument  $\{\mathcal{I}_a\}$ the three notions coincide. Indeed, in the absence of the input $x$ the decompositions of Eqs.~\eqref{eq: set compressible 1} and \eqref{eq: set simmulable 1} become identical. Moreover, in the decomposition of Eq.~\eqref{eq: set compressible 1}, instead of being revealed, the input $a$, produced by the encoder, could be merged with the hidden variable $\lambda$ and later revealed by a latent instrument. Thus the decomposition of Eq.~\eqref{eq: set embedable} is also equivalent.

\section{Special cases}

\label{sec: special cases}

Having introduced the nonequivalent definitions of dimension bottleneck for sets of quantum instruments, it is interesting to see what remains of this hierarchy for specific cases:  channels (no output $a$), measurements ($m=1$), and states ($n=1$, no output $a$). But first let us discuss the special case of $d=1$, which enforces a classical bottleneck -- all information must be compressed to classical variables.

\subsection{Trivial bottleneck dimension ($d=1$)}
\label{sec: d=1}
The special case of trivial bottleneck dimension $d=1$, where the information is compressed to a classical variable $\lambda$, deserves special attention. These decompositions can be understood as imposing a classical bottleneck on the instruments, thereby defining natural notions of non-classicality.

We know that a set of instruments $\{\mathcal{I}_{a|x}\}$ is $1$-compressible iff all CP maps $\mathcal{I}_{a|x}$ admit a Kraus decompositions with rank-one Kraus operators, i.e. iff all the instruments are {\it entanglement-breaking}~\cite{pusey2015verifying}

% . Therefore,
% \vspace{2 mm}

% \noindent
% {\it  The set $\{\mathcal{I}_{a|x}\}$ is $1$-compressible iff all the CP maps $\mathcal{I}_{a|x}$ are entanglement-breaking.}
% \vspace{2 mm}

This is also a necessary condition for $1$-simulability of the set  $\{\mathcal{I}_{a|x}\}$, however it is insufficient. To see this, using the Choi-Jamio{\l}kowski isomorphism~\cite{Jamiolkowski72}, let us write the problem of finding a $1$-simulable decomposition of the set in terms of the associated Choi operators $\{J_{\mathcal{I}_{a|x}}\}$, introduced in Def.~\ref{def: SN},
\begin{align}\label{eq: qsim prob}
    {\rm find} \quad & \{E_\lambda \succeq 0, \sigma_{a|\lambda,x}\succeq 0\} \\
    \label{eq: qsim c1}
    {\rm subject \, to} \quad & J_{\mathcal{I}_{a|x}} = \sum_\lambda E_\lambda^\top \otimes \sigma_{a|\lambda,x} \quad \forall a,x\\ % [REVIEW] was \forall x,\lambda -- lambda is summed over, and the constraint must hold for every a.
    &\sum_\lambda E_\lambda =\id ,\\
    & \tr \sum_a \sigma_{a|\lambda,x} = 1 \quad \forall x,\lambda,
\end{align}
where $\{E_\lambda\}$ is the ``parent'' POVM performed by the encoder ${\tt Enc}_\lambda[\cdot] := \tr(E_\lambda \cdot)$, and $\sigma_{a|\lambda,x}={\tt Dec}_{a|\lambda,x}$ are the subnormalized states prepared by the decoder. 
Tracing out the output system to define the probabilities $p(a|\lambda,x)= \tr \sigma_{a|\lambda,x}$, one sees that a $1$-simulable set, induces POVMs $\{M_{a|x}\}$ that are {\it jointly-measurable}~\cite{busch2016quantum,heinosaari2016invitation}
\begin{align}
    \tr (M_{a|x}\cdot):=\tr (\mathcal{I}_{a|x}[\cdot]) =  \sum_\lambda p(a|\lambda,x) \tr (E_\lambda \cdot).
\end{align}
This condition is of course not implied by $\mathcal{I}_{a|x}$ being entanglement-breaking. Additionally, if all the instruments are measure-and-prepare, i.e.~$\mathcal{I}_{a|x}[\cdot]= \tr(M_{a|x}\cdot) \, \rho_{a,x}$, joint-measurability of $\{M_{a|x}\}$ is sufficient for $1$-simulability of the instruments, since the post-measurement states $\sigma_{a|\lambda,x}=p(a|\lambda,x)\,\rho_{a,x}$ are determined by the classical variables $a$ and $x$. In general however, being entanglement-breaking and inducing jointly-measurable POVMs is not sufficient for $1$-simulability, as we will see in the examples section.
Conceptually, one can think of a $1$-simulable set as being {\it jointly-entanglement-breaking}, in the sense that the encoding of the quantum input into a classical variable is done with a fixed measurement $\{{\tt Enc}_\lambda\}$, independent of $x$.

The remaining notion of $1$-embeddable instruments is in general too restrictive to be interesting, since it removes any causal relation between the classical input $x$ and the quantum output of the instrument.

Finally, let us emphasize that our notions of 1-compressibility and 1-simulability differs from the various notions of (traditional, weak and parallel) {\it instrument compatibility} introduced in Ref.~\cite{mitra2022compatibility}. While our goal is to impose a classical bottleneck on the instruments, Ref.~\cite{mitra2022compatibility} rather demands that all instruments from the set can be realized ``simultaneously'' and recovered at the end by various $x$-dependent post-processings.  In particular, neither of three instrument compatibility notions of Ref.~\cite{mitra2022compatibility} requires the instruments to be entanglement-breaking. 

\subsection{Sets of quantum measurements}

Measurements $\{M_{a|x}\}$ are special instruments with no quantum output, i.e. a trivial output dimension $m=1$, which makes the decoder superfluous. It is therefore direct to see that for measurements the notion of $d$-compressibility is trivial, i.e. all sets of measurements are $d$-compressible. 

For the same reason,  the notions of $d$-simulability and $d$-embeddability become equivalent, and recover the notion of measurement $d$-simulability introduced earlier~\cite{ioannou2022simulability}.  As already mentioned, for $d=1$ measurement $1$-simulability is equivalent to joint measurability.

\subsection{Sets of quantum states}

\begin{figure}[t!]
    \centering
    \includegraphics[height=2.4cm]{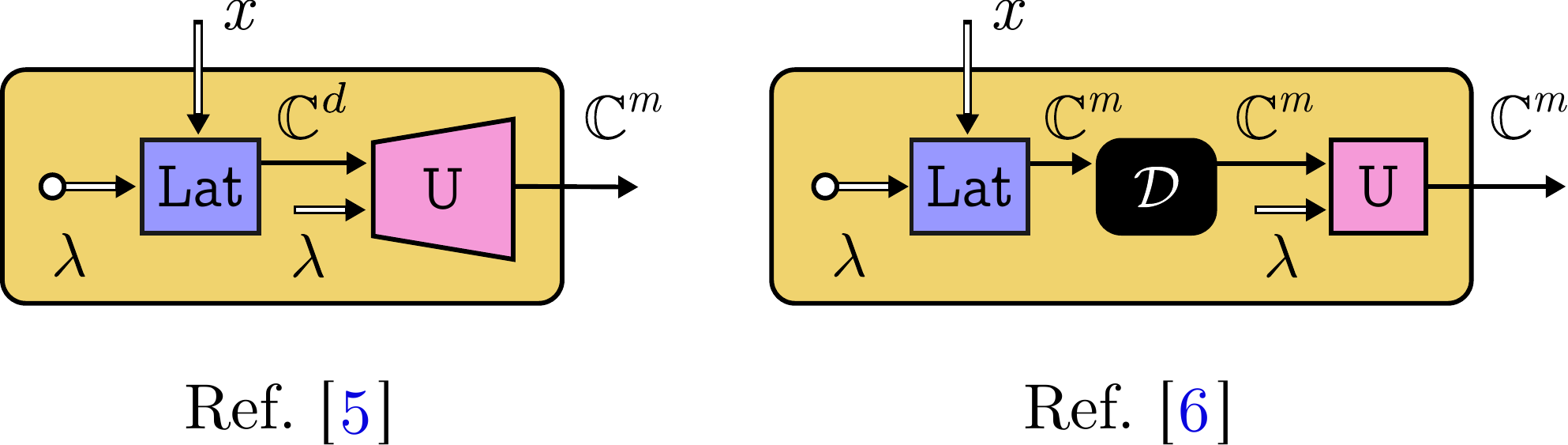}
    \caption{{(\bf Left)} Interpretation  of the notion of ``absolute dimensionality'' of a set of quantum states $\{\rho_x\}$~\cite{bernal2024absolute} from the dimension bottleneck perspective. Here $\{{\tt U}_\lambda\}$ are decoder channels that are restricted to perform isometric embeddings from dimension $d$ to $m$.
    {(\bf Right)} Interpretation of the notion of ``operationally classical simulation'' of a set of quantum states $\{\rho_x\}$~\cite{cobucci2026operationally}. Here $\mathcal{D}$ is the dephasing channel (e.g. in the computational basis), such that all states received by the unitary decoder $\{{\tt U}_\lambda\}$ are diagonal in the same basis.}
    \label{fig: armin} 
\end{figure}

Quantum states $\{\rho_x\}$ are special instruments with no quantum input ($n=1$) and no classical output $a$. Due to the absence of a quantum input any set of states is trivially $1$-compressible and $1$-simulable. In turn, the $d$-embeddability condition for a set of quantum states reads
\be\label{eq:1-e states}
d\textit{-embeddable}:\quad \rho_x = \sum_\lambda  p(\lambda) {\tt Dec}_\lambda[\sigma_{\lambda,x}]
\ee
where $\sigma_{\lambda,x} := {\tt Lat}_{|\lambda,x}$, and the decoder can perform  any maps from dimension $d$  to $m$. This notion is closely related to that of ``absolute dimensionality'' in Ref.~\cite{bernal2024absolute} (see Fig.~\ref{fig: armin}), which  requires the existence of a decomposition
\be\label{eq: armin}
\textit{absolute dimension~\cite{bernal2024absolute}}:\quad \rho_x = \sum_\lambda  p(\lambda) {\tt U}_\lambda[\sigma_{\lambda,x}]
\ee
where ${\tt U}_\lambda$ are {\it isometric embeddings} of the $d$-dimensional states $\sigma_{\lambda,x}$ in dimension $m$. 
We immediately see that this is a particular case of the above decomposition, i.e. Eq.~\eqref{eq: armin} $\implies$ Eq.~\eqref{eq:1-e states}, whereas it is at present unclear if the inverse implication holds in general. In any case, we conclude that the ``absolute dimensionality'' of a set of states introduced in Ref.~\cite{bernal2024absolute} can be understood via the notion of $d$-embeddability with the decoder restricted to perform isometric embeddings.

Since any set of states is 1-simulable and the notion of 1-embeddability is uninteresting, one is left wondering how an operational, basis-independent, notion of classicality could be defined for quantum states. A way to do so has been proposed in Ref.~\cite{cobucci2026operationally}. In our language this definition can be understood as follows (Fig.~\ref{fig: armin}): one allows the latent instrument to send states that, instead of being bounded in dimension, must be {\it diagonal} in the computational basis.

% Thus $1$-embeddability does not give an interesting notion of classicality. For quantum states a way around this limitation has been proposed in Ref.~\cite{bernal2024absolute} to define the notion of "classicality" for sets of states. In our language this can be understood as follows (see Fig), one  allows the latent instrument to send states that instead of being bounded in dimension  must be {\it diagonal} in the computational basis. Recently, a similar idea has also been applied to measuremtns~[].

%%%%%%%%%%%%%%%%%%%%%%%%%%%%%%%%%%%%%%%%%%%%%%%%%%%
%%%%%%%%%%%%%%%%%%%%%%%%%%%%%%%%%%%%%%%%%%%%%%%%%%%
\section{Examples}
\label{sec: examples}
To illustrate the introduced concepts we now discuss a series of examples: complementary qubit measurements, the white noise channel, and the complete set of noisy unitary channels.

\subsection{The usual suspects}

\begin{table*}[t] % [REVIEW] was \begin{table}[H]
\centering
\begin{tabular}{|c||c|c|}
\hline
& 1-compressible & 1-simulable
\\
\hline\hline
Ideal weak measurements (L\"uder's instruments)
$\{\mathcal{I}_{a|x}^{(p)}\}$ in Eq.~\eqref{eq: XZ ideal}& iff $p=1$ & never \\
\hline
Ideal projective measurements with noisy classical output
$\{\mathcal{P}_{a|x}^{(p)}\}$ in Eq.~\eqref{eq: proj  noisy} & always& never\\
\hline
Measure-and-prepare instruments 
$\{\mathcal{M}_{a|x}^{(p)}\}$ in Eq.~\eqref{eq: MP instruments} & always& iff $p\leq \frac{2+\sqrt 2}{4}$ \\
\hline
\end{tabular}
\caption{Classical simulability of different instruments realizing the noisy two-outcome qubit POVMs $\{ M_{a|x}^{(p)}=  p\, \Pi_{a|x}+(1-p)(\id-\Pi_{a|x})\}$ in the $Z$ and $X$ bases, in function of the parameter $p\in[\nicefrac{1}{2},1]$. Recall that the notion of 1-embeddability is trivial (Sec.~\ref{sec: d=1}) and thus not reported here.}
\label{tab: tab1}
\end{table*} % [REVIEW] was \end{table}

As the first example, we consider the paradigmatic noisy qubit measurements in the $Z$ and $X$ bases, given by the POVM elements
\begin{align}\nonumber
M_{0|0}^{(p)}&= p \, \Pi_0 +(1-p) \Pi_1 &\,  M_{1|0}^{(p)}&= p \, \Pi_1 +(1-p) \Pi_0\\
M_{0|1}^{(p)}&= p \, \Pi_+ +(1-p) \Pi_- &\,  M_{1|1}^{(p)}&= p \, \Pi_- +(1-p) \Pi_+ \nonumber,
\end{align}
where $\Pi_{x}=\ketbra{x}$ are projectors on the respective qubit states; for which we will use the compact  notation $\Pi_{0|0}= \Pi_0,   \Pi_{1|0}=  \Pi_1, \Pi_{0|1}=  \Pi_+,$ and $\Pi_{1|1}=  \Pi_-.$ The set $\{M_{a|x}^{(p)}\}$ is known to be jointly-measurable iff $p\leq \frac{2+\sqrt 2}{4}$~\cite{busch1986unsharp}. The optimal parent POVM has four elements
\be
E_{ij} = \frac{1}{4} \left(\id +\frac{(-1)^i\sigma_Z +(-1)^j\sigma_X}{\sqrt{2}}\right)
\ee
with $i,j\in \{0,1\}$, proportional to the four projectors equally close to the X and Z directions in the Bloch sphere. The post-processing is given by $M_{a|x}^{(p)}= \sum_{ij} p(a|ij,x) E_{ij}$ with 
\be
p(a|ij,x) =
\begin{cases}
\delta_{i a} &\text{if } x=0\\
\delta_{j a} &\text{if } x=1
\end{cases}.
\ee
\noindent This reproduces $M^{(p)}_{a|x}$ exactly at the critical value $p=\frac{2+\sqrt2}{4}$; for $p$ below it, one simply mixes the post-processing with white noise. 
The measurements $\{M_{a|x}^{(p)}\}$ do not have a quantum output, which physically means that during the measurement process the measured system is either discarded or reinitialized in a deterministic state (demolished).There are, however, other instruments inducing the same paradigmatic measurements upon discarding the post-measurement state.  We now consider three such sets of instruments, and discuss when those become $1$-compressible and $1$-simulable. Note that by symmetry we can restrict the noise parameter to the interval $p\in[\nicefrac{1}{2},1]$, the case $p<\nicefrac{1}{2}$ can be recovered by a relabeling of the output $a$.\\

As the first possibility, consider the ideal weak measurements in the $X$ and $Z$ bases. They are given by the L\"uder's instruments~\cite{luders1950,*luders2006translation} of our POVMs
\begin{align}\label{eq: XZ ideal}
\mathcal{I}^{(p)}_{a|x}[\cdot] &= \sqrt{M^{(p)}_{a|x}}\cdot \sqrt{M^{(p)}_{a|x}}.
\end{align}
 For $\nicefrac{1}{2} \leq p <1$, the CP maps $\mathcal{I}^{(p)}_{a|x}[\cdot]$ are not entanglement-breaking. Hence, in this parameter range the set $\{\mathcal{I}^{(p)}_{a|x}\}$ is neither $1$-compressible nor $1$-simulable. For $p=1$, these are ideal projective measurements which are $1$-compressible but not $1$-simulable.\\

Second, consider the instruments 
\begin{align}\label{eq: proj  noisy}
    \mathcal{P}^{(p)}_{a|x}[\cdot] &= p \,\Pi_{a|x} \cdot \Pi_{a|x} + (1-p) \Pi_{a\oplus 1|x} \cdot \Pi_{a\oplus 1|x},
\end{align}
which describe an ideal projective measurement followed by bit-flip noise on the output $ \mathcal{P}^{(p)}_{a|x} = p \,  \mathcal{I}^{(1)}_{a|x}+  (1-p)\, \mathcal{I}^{(1)}_{a\oplus 1|x}$. These instruments are entanglement-breaking and thus $1$-compressible. In contrast, the set $\{\mathcal{P}^{(p)}_{a|x}\}$ is never $1$-simulable. This can be seen by considering the worst case scenario $p=\nicefrac{1}{2}$, where the instruments $\mathcal{P}^{(\nicefrac{1}{2})}_{0|x}=\mathcal{P}^{(\nicefrac{1}{2})}_{1|x} =\frac{1}{2} \mathcal{D}_x$ are equivalent to the nonselective measurement (dephasing) channels $\mathcal{D}_x[\cdot] = \sum_a \Pi_{a|x} \tr(\Pi_{a|x} \cdot)$, with the Choi operators
\begin{align}
J_{\mathcal{D}_0} &= \frac{1}{2}\left( \Pi_0\otimes \Pi_0 + \Pi_1\otimes \Pi_1\right)\\
J_{\mathcal{D}_1} &= \frac{1}{2}\left( \Pi_+\otimes \Pi_+ + \Pi_-\otimes \Pi_-\right).
\end{align}
Hence for each element $E_\lambda$ of the parent POVM, Eq.~\eqref{eq: qsim c1} implies that the two constraints
\begin{align}
J_{\mathcal{D}_0}  &\succeq E_\lambda^\top \otimes \sigma  \qquad 
J_{\mathcal{D}_1} \succeq E_\lambda^\top \otimes \sigma'  
\end{align}
must hold simultaneously for some nonzero $\sigma,\sigma'\succeq 0$. 
The first constraint forces ${\rm supp}(E_\lambda^\top\otimes \sigma)\subseteq {\rm span}\{\ket{00},\ket{11}\}$ and the second ${\rm supp}(E_\lambda^\top\otimes \sigma')\subseteq {\rm span}\{\ket{++},\ket{--}\}$. Since $E_\lambda^\top\otimes\sigma$ is a product operator, its support is spanned by product vectors; but the only product vectors in the first subspace are $\ket{00}$ and $\ket{11}$, and in the second $\ket{++}$ and $\ket{--}$. These two families are disjoint, so $E_\lambda=0$ for every $\lambda$, contradicting $\sum_\lambda E_\lambda=\id$.\\

Finally, consider the instruments
\begin{align}\label{eq: MP instruments}
    \mathcal{M}^{(p)}_{a|x}[\cdot] &= \Pi_{a|x}\, \tr(  M_{a|x}^{(p)}\cdot),
\end{align}
which perform the paradigmatic measurements $\{M^{(p)}_{a|x}\}$, and then prepare the ideal post-measurement state determined by $x$ and $a$. These instruments are always entanglement-breaking and thus $1$-compressible. In addition they are by construction measure-and-prepare, and therefore $1$-simulable iff the induced POVMs are jointly-measurable, i.e.~iff $p\leq \frac{2+\sqrt 2}{4}$. To see this explicitly, consider the following decomposition 
\be
\mathcal{M}^{(p)}_{a|x}[\cdot] = \sum_{ij} \sigma_{a|ij,x} \tr(  E_{ij}\cdot),
\ee
with $\sigma_{a|ij,x} = p(a|ij,x) \Pi_{a|x}$.\\

These three examples, summarized in  Table~\ref{tab: tab1}, illustrate the subtle interplay between the $1$-compressibility and $1$-simulability of the instruments and the properties of the post-measurement states. We see that the joint-measurability of the induced POVMs, only gives a very coarse-grained picture of the ``classical simulability'' of the initial instruments.

\subsection{The white noise channel}

As our second example we consider the white-noise, also known as partially-depolarizing, channel in dimension $n$
\be
\mathcal{W}_p [\rho] := p \rho + (1-p)\frac{\id}{n}\tr(\rho).
\ee
Due to its symmetry,  the white noise is often taken as a generic source of imperfections in theoretical quantum information.
The Schmidt number of the white noise channel is known~\cite{terhal2000schmidt}
\be
{\rm SN}(\mathcal{W}_p) = \left \lceil n (p + \frac{1-p}{n^2}) \right\rceil,
\ee
hence it is $d$-compressible iff $p\leq \frac{d\, n -1}{n^2-1}$.\\

% For the following, it is also insightful to look at an explicit $d$-compressible decomposition (Fig.~\ref{fig: generalchan}), exploiting the symmetry of $\mathcal{W}_p $. Specifically, consider the following CPTP map
% \be\label{eq: WN dec}
% \mathcal{P}_d[\cdot]=\! \int \!\dd V \! \left({\tt Dec}_{|V} \circ {\tt Enc}_V\right) \![\cdot] = \!\int \!\frac{n}{d}\dd V\,   \Pi_V^{(d)} \cdot \Pi_V^{(d)},
% \ee
% where $ \Pi^{(d)}= \sum_{i=1}^d \ketbra{i}$ is a projector on a fixed $d$-dimensional subspace and  $\dd V$ is the Haar measure over the unitaries in dimension $n$, such that $\Pi_V^{(d)} := V \Pi^{(d)} V^\dag$ are projectors on Haar-random subspaces. The decomposition combines the instument $\{{\tt Enc}_V\}$ with Kraus operators (density) $E_V = \sqrt{\frac{n}{d}}\Pi^{(d)} V$ and a continuous-valued classical output $\lambda = V$, and the continuous set of decoding channels $\{{\tt Dec}_{|V}\}$ performing the isometric embedding ${\tt Dec}_{|V}[\cdot]= V \Pi^{(d)}\cdot \Pi^{(d)} V^\dag$ back in dimension $n$.  In section ..., we explicitly demonstrate that 
% \be
% \mathcal{P}_d=\mathcal{W}_{\frac{n\, d-1}{n^2-1}},
% \ee
% our scheme thus the optimal $d$-compression of the white noise channel.

Given the prevalence of white noise in theoretical quantum information, it is interesting to see how it affects the compressibility of other sets of operations. More generally, the following proposition clarifies the impact of noise on this property.
\begin{proposition}
For a set of instruments  $\{\mathcal{I}_{a|x}\}$, and a $d$-compressible channel $\cE$,  the set of noisy instruments $\{\tilde{\mathcal{I}}_{a|x} = \mathcal{I}_{a|x}\circ \cE\}$ is $d$-simulable.
\end{proposition}

\noindent The proof is single-line.  Given the decomposition $\cE = \sum_\lambda {\tt Dec}_{|\lambda} \circ {\tt Enc}_{\lambda} $, simply define $\tilde {\mathcal{I}}_{a|x}= \sum_\lambda {\tt Dec}'_{a|\lambda,x} \circ {\tt Enc}_{\lambda}$ with ${\tt Dec}'_{a|\lambda,x} := \mathcal{I}_{a|x}\circ {\tt Dec}_{|\lambda}$.  
For this decomposition to hold it is crucial that the noise channel is applied before the instrument. For instance, $d$-simulability of the set $\{\overline{\mathcal{I}}_{a|x} = \cE \circ \mathcal{I}_{a|x}\}$ is not a priori guaranteed.\\

Applying the proposition to white noise gives an immediate corollary. 
\begin{corollary}\label{cor: WP}
Any noisy set of instruments  $\{\tilde{\mathcal{I}}_{a|x} = \mathcal{I}_{a|x}\circ \mathcal{W}_p\}$ is $d$-simulable if $p\leq \frac{d\, n-1}{n^2 -1}$. If the set contains a unitary channel $\mathcal{U} \circ\mathcal{W}_p$, it is not $d$-compressible for $p > \frac{d\, n-1}{n^2 -1}$\new{.}
\end{corollary}

The converse direction follows from the fact that $\mathcal{U} \circ\mathcal{W}_p$ is only $d$-compressible if $\mathcal{W}_p$ is, as the unitary can always be absorbed in the decoder.

\subsection{The set of all noisy unitaries}
\label{sec: all U}

As our last example consider the set of {\it all noisy unitary} operations 
\be
\{\mathcal{U}^{(p)} := \mathcal{U}\circ \mathcal{W}_p\}
\quad \text{with} \quad \mathcal{U}[\cdot]= U \cdot U^\dag,
\ee
where $U$ runs through the whole special unitary group in dimension $n$. In this case the order in which the noise and the unitary channels act is irrelevant, and each noisy unitary channel 
\be
\mathcal{U}^{(p)} = p\,  \mathcal{U} + (1-p) \mathcal{R}_{\nicefrac{\id}{n}}
\ee
is a convex combination of the ideal unitary and a replacement (aka trash-and-replace) channel $ \mathcal{R}_{\nicefrac{\id}{n}}[\cdot] = \tr[\cdot]\,  \nicefrac{\id}{n}$. It will be useful to compute the Choi--Jamio{\l}kowski fidelity of the noisy with the ideal unitary channels \cite{raginsky2001fidelity}. 
%,tripathi2025benchmarking
 As $J_{\mathcal{U}}/n$ is a pure state, the fidelity is given by
\begin{align}\label{eq: F Up}
F_{\rm CJ}^2(\mathcal{U}^{(p)},\mathcal{U})&=
%\tr \Big(({\rm id}\otimes \mathcal{U})[\Phi^+] ({\rm id}\otimes \mathcal{U}^{(p)})[\Phi^+] \Big)\\
\tr \left(\frac{ J_{\mathcal{U}} }{n} \frac{J_{\mathcal{U}^{(p)}}}{n}\right)= p + \frac{(1-p)}{n^2}.
\end{align}
By the Corollary~\ref{cor: WP} the set $\{\mathcal{U}^{(p)}\}$ is $d$-compressible and $d$-simulable if and only if  $p\leq \frac{d\, n-1}{n^2 -1}$. \\

It is more challenging to figure out when $\{\mathcal{U}^{(p)}\}$ is $d$-embeddable. For this purpose it is helpful to note that our set is {\it input-output-invariant}, i.e.~for any $\mathcal{U}^{(p)}$ and any unitary channels $\mathcal{V}_i$ and $\mathcal{V}_o$ the composed channel $\mathcal{V}_o \circ\mathcal{U}^{(p)}\circ\mathcal{V}_i$ remains within the original set. In App.~\ref{sec:proof invariance} we show that for such highly symmetric sets of channels (or instruments), the encoding and decoding instruments can also be assumed highly-symmetric without loss of generality. This allows us to only look for decompositions  of the form
 \begin{align} \nonumber
   \mathcal{U}^{(p)} \overset{?}{=}\cE_{U}= \!\int \!\dd V_i \dd V_o \, {\tt Dec}_{|V_{i},V_{o}}  \circ {\tt Lat}_{|V_{i},V_{o},U} \circ {\tt Enc}_{V_{i},V_{o}},
\end{align}
with the following continuous-valued operations
\begin{align}
{\tt Enc}_{V_{i},V_{o}} [\cdot] &= \frac{n}{d}   \Pi^{(d)} \, V_{i} \cdot V_{i}^\dag \Pi^{(d)}\\
{\tt Dec}_{|V_{i},V_{o}} [\cdot] &=  V_{o}\Pi^{(d)} \cdot \Pi^{(d)} V_{o}^\dag,
 \end{align}
where $V_i$ and $V_o$ are unitary matrices from $SU(n)$ sampled from the Haar measures $\dd V_{i}$ and $\dd V_{o}$, and the classical variable $\lambda=(V_i,V_o)$ contains the classical description of these matrices. Moreover, we show in App.~\ref{sec: optimality of cEU} that the latent channels can also be assumed unitary ${\tt Lat}_{|V_i,V_o,U}[\cdot] := W_{V_i,V_o,U} \cdot W_{V_i,V_o,U}^\dag$. Let us now compute the Choi--Jamio{\l}kowski fidelity between $\cE_{U}$ and $\mathcal{U}$
\begin{align}\nonumber
    &F_{\rm CJ}^2(\cE_U,\mathcal{U})=
%\tr \Big(({\rm id}\otimes \mathcal{U})[\Phi^+] ({\rm id}\otimes \mathcal{U}^{(p)})[\Phi^+] \Big)\\
\tr \left(\frac{ J_{\mathcal{U}} }{n} \frac{J_{\cE_U}}{n}\right)\\
&=\! \int \!\dd V_i \dd V_o \frac{1}{n^2}\left|\tr (U^\dag V_o\Pi^{(d)}W_{V_i,V_o,U} \sqrt{\frac{n}{d}} \Pi^{(d)}V_i )\right|^2\nonumber
\\
%&=\frac{1}{d n}\! \int \!\dd V_i \dd V_o \left|\tr (W_{V_i,V_o,U} \Pi^{(d)}V_i U^\dag V_o\Pi^{(d)})\right|^2\nonumber\\
&\leq \frac{1}{ d n}\! \int \!\dd V_i \dd V_o \left \| \Pi^{(d)}V_i U^\dag V_o\Pi^{(d)}\right\|_*^2 \nonumber\\
& =\frac{d}{n} \int \dd V \frac{1}{d^2}\| \Pi^{(d)} V \Pi^{(d)}\|_*^2 =: \frac{d}{n} \, \mathfrak{f}_2(d,n), \label{eq: F EC}
\end{align}
where $\|\cdot\|_*$ is the nuclear norm (sum of singular values). The inequality in the fourth line is saturated by choosing the latent unitary $W_{V_i,V_o,U}$ dictated by the polar decomposition, i.e.~such that 
\be
W_{V_i,V_o,U} \Pi^{(d)}V_i U^\dag V_o\Pi^{(d)}:= |\Pi^{(d)}V_i U^\dag V_o\Pi^{(d)}|.
\ee
Our choice of ${\tt Enc}_{V_{i},V_{o}}$, ${\tt Dec}_{|V_{i},V_{o}} $ and ${\tt Lat}_{|V_i,V_o,U}$ thus gives the {\it optimal embedding strategy} for the set of noisy unitary channels.

For completeness, let us verify that the constructed channel $\cE_U$ has the desired form $\cE_U = \mathcal{U}^{(q)}$ for some value of the parameter $q$. Put differently, we verify that 
\be\label{eq: rhs}
\mathcal{C}_U:=\mathcal{U}^{-1} \circ \cE_U = \mathcal{W}_{q}
\ee
for some $q$.  To do so, notice that by the Schur lemma~\cite{nechita2026random} any channel $\mathcal{C}$ satisfying $\mathcal{C} = \mathcal{V}\circ\mathcal{C}\circ \mathcal{V}^\dag$ for all unitary channels $\mathcal{V}$ must be a depolarizing noise channel $\mathcal{W}_q$. As our construction of $\cE_U$ is based on the Haar measure, it is not difficult to see (App.~\ref{sec: CU inv}) that $\mathcal{C}_U$ has the desired symmetry, i.e.~$\mathcal{V}\circ\mathcal{C}_U\circ \mathcal{V}^\dag = \mathcal{C}_U = \mathcal{C}_\id$. Therefore, $\cE_U = \mathcal{U}^{(q)}$, and the value of $q$ can be computed by combining the Eqs.~\eqref{eq: F EC} with \eqref{eq: F Up}. 

\begin{proposition} The set of all noisy unitary channels $\{\mathcal{U}^{(p)}:= \mathcal{U}\circ \mathcal{W}_p\}$ in dimension $n$ is $d$-compressible and $d$-simulable iff 
\be
p\leq p_{\rm S}^*(d,n):=\frac{d\, n -1}{n^2 -1}\approx \frac{d}{n},
\ee 
and $d$-embeddable iff
\be\label{eq: emb Up}
p\leq p_{\rm E}^*(d,n) :=\frac{d\,n\, \mathfrak{f}_2(d,n) -1}{n^2 -1}{\color{blue}.}
\ee 
\end{proposition}
\noindent
 Numerical values of the white noise thresholds for different values of $n$ and $d$ are presented in Table~\ref{tab: tab2}.

\begin{table}[t] % [REVIEW] was \begin{table}[H]
\centering
\begin{tabular}{|c||cc|cc|cc|cc|cc|cc|cc|}
\hline
\diagbox{$n$}{$d$} & \multicolumn{2}{c|}{1}& \multicolumn{2}{c|}{2} & \multicolumn{2}{c|}{3} & \multicolumn{2}{c|}{4}
& \multicolumn{2}{c|}{5} & \multicolumn{2}{c|}{6} & \multicolumn{2}{c|}{7} 
\\
\hline
 & $p_{\rm S}^*$&$p_{\rm E}^*$ &$p_{\rm S}^*$&$p_{\rm E}^*$ & $p_{\rm S}^*$&$p_{\rm E}^*$& $p_{\rm S}^*$&$p_{\rm E}^*$&  $p_{\rm S}^*$&$p_{\rm E}^*$& $p_{\rm S}^*$&$p_{\rm E}^*$& $p_{\rm S}^*$&$p_{\rm E}^*$
\\
\hline\hline
2 &33 & 0& & & & & & & & & & &  &
\\
3 &25&0& 63&33& & & & & & & & &  &
\\
4 &20&0& 47&16& 73&47 & & & & & & & & 
\\
5 &17&0& 38&10&  58&28&   79&56& & & & &&  
\\
6 &14&0& 31&7& 49&19 & 66&37 & 83&63 & & &&  
\\
7 &13&0& 27&5& 42&13 & 56&26 & 71&44 & 85&67 &&  
\\
8 &11& 0&  24&4& 37&10 & 49&19 & 62&32  &  75&49  & 87&71 
\\
\hline
\end{tabular}
\caption{ The critical values $p$ (in percent) below which the set of all unitary channels combined with white noise $\{\mathcal{U}^{(p)}:= \mathcal{U}\circ \mathcal{W}_p\}$ in nominal dimension $n$ become $d$-simulable ($p_{\rm S}^*$ left value) and $d$-embeddable ($p_{\rm E}^*$ right value).}
\label{tab: tab2}
\end{table} % [REVIEW] was \end{table}

In the above derivation we have introduced the quantity $\mathfrak{f}_2(d,n):= \frac{1}{d^2}\mathds{E}_{V\sim {\rm CUE}(n)} \left[\| \Pi^{(d)} V \Pi^{(d)}\|_*^2\right]$, describing a specific property of the circular unitary ensemble (CUE) of the Haar-random unitary matrices\footnote{For our discussion it does not matter if one considers the whole unitary ensemble (CUE) of matrices $V$ or the special unitary restrictions}. This quantity is more natural to view as a property $\mathfrak{f}_2(d,n):= \frac{1}{d^2}\mathds{E}_{A\sim {\rm JUE}(d,n)} \left[\|A\|_*^2\right]$ of the Jacobi unitary ensemble~\cite{zyczkowski2000truncations,potters2020first}, also known as truncated unitary ensemble, obtained by restricting random $n\times n$ unitary matrices to a fixed block of size $d\times d$. One immediately sees that 
\be
\frac{1}{d^2}\mathds{E} \left[\|A\|_*^2\right] = \frac{1}{d} \,\mathds{E} [\lambda_1^2] + \frac{d-1}{d}\mathds{E} [\lambda_1\lambda_2],
\ee
where $\lambda_1$ and $\lambda_2$ are two distinct singular values of $A\sim {\rm JUE}(d,n)$. Note that $\mathds{E}[\lambda_1^2]=\nicefrac{d}{n}$ is known exactly, since $\mathds{E}[\tr A^\dag A]=d^2/n$, so that only the cross-moment $\mathds{E}[\lambda_1\lambda_2]$ is missing.
While ${\rm JUE}(d,n)$ is a standard and well-studied ensemble in random matrix theory, there seems to be no known general closed-form expression of $\mathds{E} [\lambda_1\lambda_2]$. We have thus computed the values in Table~\ref{tab: tab1} numerically, by sampling from the CUE, and only derived the analytic expression of $\mathds{E} [\lambda_1\lambda_2]$ for the special case $d=n-1$ in App.~\ref{app: f2 top}.

The ideal set of all unitary channels we considered describes a universal quantum processor, capable to implement any invertible transformation of a quantum system. We have seen that in presence of  depolarizing noise, such a processor quickly loses the global coherence across the full $n$-dimensional Hilbert space. More precisely, for $d=n-1$ we find
\begin{align}
    p_{\rm S}^*(n-1,n) &= 1-\frac{1}{n}+ O(n^{-3})\\
    p_{\rm E}^*(n-1,n) &= 1-\frac{3}{n} +O(n^{-3/2}),
\end{align}
where the last value is computed in App.~\ref{app: f2 top}. Preserving the quantum coherence across the whole Hilbert space is thus increasingly challenging for larger quantum processors. 

On the other hand, from the bound on $d$-simulability we see that it is the ratio $\nicefrac{d}{n}$ of the bottleneck and nominal dimensions that determines the white noise threshold. Thus, in terms of global white noise, realizing a processor with a large quantum bottleneck dimension $d$ is easier for a system of large nominal dimension $n$. However, in practice this optimistic conclusion is to be taken with a grain of salt for two reasons. First, the global white-noise model is overly simplistic. Second, the strength of the noise is in general not independent from the dimension of the Hilbert space. For instance, the number of elementary (noisy) single and two-qubit gates required to approximate a universal quantum processor increases quadratically with the dimension $n$~\cite{mottonen2004quantum, nielsen2010quantum}.

% \begin{proposition}
% An input-output-invariant set of instruments $\{\mathcal{I}_{a|x}\}$, with $\mathcal{I}_{a|x} : L(\mathds{C}^n)\to L(\mathds{C}^m) $,  is $d$-embeddable iff it is $d$-embeddable with the following continuously-valued encoding and decoding operations
% \begin{align}
% {\tt Enc}_{V_{i},V_{o}} [\cdot] &= \frac{n}{d}   \Pi^{(d)} \, V_{i} \cdot V_{i}^\dag \Pi^{(d)}\\
% {\tt Dec}_{|V_{i},V_{o}} [\cdot] &=  V_{o}\Pi^{(d)} \cdot \Pi^{(d)} V_{o}^\dag.
%  \end{align}
% with respect to the Haar measure $\dd V_{i}$ and $\dd V_{o}$ on $SU(n)$ and $SU(m)$\footonote{Notice the slight abuse of notation }.
% \end{proposition}
% The proof is postponed to Sec.~\ref{sec:proof invariance}.

%%%%%%%%%%%%%%%%%%%%%%%%%%%%%%%%%%%%%%%%%%%%%%%%%%%
%%%%%%%%%%%%%%%%%%%%%%%%%%%%%%%%%%%%%%%%%%%%%%%%%%%
\section{Discussion }

Before the closure, it is worth abstracting away from the mathematical details and revisiting our three definitions of the dimension bottleneck of a set of operations $\{\mathcal{I}_{a|x}\}$ from a more physical or information theoretic perspective. 

\subsection{Quantum communication with limited dimension perspective}

The notions of compressibility and simulability are best understood through a quantum communication scenario. Concretely, let us assume that the quantum input of the instrument is sent by Alice, and its quantum output is received by Bob, who are separated in space. We then ask for the minimal dimension of the quantum system that must be transmitted from Alice to Bob in order to realize the instruments $\{\mathcal{I}_{a|x}\}$.

The  notion of  $d$-compressibility corresponds to the case where Alice knows  which instrument from the set must be realized from the beginning. Hence, it does not depend on the relation between different instruments in the set, and is directly given by their maximal Schmidt number. Indeed, Alice can simply perform the desired instrument and send a compressed quantum output to Bob, alongside $x$, the  classical output $a$ and auxiliary classical variables. The required transmitted dimension is then given by the maximal Schmidt number among all CP maps $\mathcal{I}_{a|x}$.

The case of $d$-simulability is more subtle. Here, Alice does not know which input $x$ will be selected by Bob. Hence, her strategy and the quantum communication protocol must allow Bob to simulate all instruments in the set. He receives the input $x$ and is free to do arbitrary post-processing of the received quantum systems and classical variables. Whenever the set contains more than one instrument, this task is intuitively harder; and we have indeed seen that there is a strict separation between $d$-compressible and $d$-simulable sets of instruments, channels and measurements.\\

\subsection{Action-in-low-dimension perspective}

In principle, the decompositions with a latent instrument ($d$-embeddability and $d$-simulability) can also be viewed operationally through the lens of a quantum communication protocol, by introducing a third party between Alice and Bob, who acts on the transmitted system. Nevertheless, here this perspective  does not seem to be the most natural one.

Let us instead focus on the existence of the latent instruments $\{{\tt Lat}_{a|\lambda,x}\}$ acting in a lower dimension $d$, which are somehow embedded in the larger Hilbert space in order to simulate $\{\mathcal{I}_{a|x}\}$. 
This embedding is done with a quantum comb, which we only allow to have a classical memory in order to impose a strict dimension bottleneck. 
Depending on how classical information exchange is organized between the comb and the latent instruments, different notions of dimension bottleneck are obtained. In particular, if the comb is allowed to send and receive a classical variable from the instrument the notion of $d$-simulability is recovered. While $d$-embeddability is recovered when the comb can send, but not receive, a classical variable. In fact, two alternative patterns are possible and lead to nonequivalent notions, see the discussion in App.~\ref{app: d-imbed}. The conceptual point here is to see that $d$-simulable or $d$-embeddable sets $\{\mathcal{I}_{a|x}\}$ can be viewed as a $d$-dimensional processor that is merely embedded in the Hilbert space of nominal dimension $n$.

Operationally, one may picture a client who delegates a programmable operation $\{\mathcal{I}_{a|x}\}$ to a service provider. The provider advertises an $n$-dimensional device, but a $d$-simulable (resp.\ $d$-embeddable) set is one that an honest-looking provider could realize while only ever maintaining coherence over  dimension $d$, padding the rest with classically correlated encoding and decoding. The bottleneck dimension is then exactly the quantum resource the client is really paying for.

%%%%%%%%%%%%%%%%%%%%%%%%%%%%%%%%%%%%%%%%%%%%%%%%%%%
%%%%%%%%%%%%%%%%%%%%%%%%%%%%%%%%%%%%%%%%%%%%%%%%%%%
\section{Conclusion and outlook}

Motivated by the intuitive idea of a dimensional bottleneck, we introduced operational and basis-independent notions aimed at quantifying the ``genuine dimension'' of programmable quantum operations. We identified three distinct notions: namely $d$-compressibility, $d$-simulability, and $d$-embeddability, corresponding to different ways in which the dimensional bottleneck may be imposed. These notions form a strict hierarchy, with $d$-embeddability being the most restrictive.  Beyond introducing new conceptual tools, the bottleneck-dimension framework also provides a unified perspective on several existing notions, including the Schmidt number of completely positive maps~\cite{huang2006schmidt}, joint measurability~\cite{busch2016quantum,heinosaari2016invitation} and measurement simulability~\cite{ioannou2022simulability} of quantum measurements, as well as the absolute dimension of quantum states~\cite{bernal2024absolute}.

To illustrate the introduced concepts we discussed in Sec.~\ref{sec: examples} a couple of examples. First, motivated by the paradigmatic example from joint-measurability, we considered three different instruments describing noisy qubit measurements in the $X$ and the $Z$ bases. We have seen that the form of the post-measurement state plays a crucial role in determining the characteristics of the instrument set, highlighting a subtler structure, which disappears when only looking at the joint-measurability of POVMs induced by the instruments. Second, we looked at the set of all noisy unitary channels in dimension $n$, which describes a universal, but noisy, quantum processor. For the simple global white noise model with visibility $p$, we have established analytical expressions of the threshold visibilities at which the set becomes $d$-compressible, $d$-simulable and $d$-embeddable. These values confirm the intuition that realizing operations that are ``genuinely quantum'' across the full Hilbert space becomes increasingly challenging for larger systems: full coherence requires a visibility above $p\geq 1-O(\nicefrac{1}{n})$.

An important direction opened by the present work is the development of methods to compute the bottleneck dimension for concrete sets of quantum operations and to witness it experimentally. For the example of all noisy unitaries, we heavily relied on the symmetry of the sets to analytically compute the threshold level of noise. This approach is clearly unsuitable for a generic set of instruments. Nevertheless, for a discrete set 
Carath\'eodory's theorem guarantees that the cardinality of the hidden variable $\lambda$ remains finite in all the decompositions. Thus, one should be able to cast the questions of $d$-compressibility, simulability or embeddability of such a set as a finite convex (but nonlinear) optimization problems. For instance,
characterizing $d$-compressibility of a set is equivalent to independently computing the Schmidt number of all instruments, and various methods for Schmidt number characterization have been proposed in the literature, see e.g.~\cite{hulpke2004simplifying,de2023complete,
PhysRevLett.134.090802,nvx7-1727}.  In turn, checking $d$-simulability and $d$-embeddability, which depend on the relations of different instruments in the set, requires the development of novel methods. We leave these questions for future work.

\begin{acknowledgments}
I am grateful to Roope Uola, Nicolas Brunner, Jef Pauwels, Sophie  Egelhaaf and Armin Tavakoli for enlightening discussion and encouraging feedback on some of the results presented in this manuscript. I acknowledge the use of AI (Claude, Anthropic) for language editing, literature search, and some assistance with the mathematics: the computation in App.~\ref{app: f2 top}, while elementary, was initially laid out by the AI and subsequently verified by me. This work was supported by the  Swiss State Secretariat for Education, Research and Innovation (SERI) under contract number UeM019-3.
\end{acknowledgments}

\bibliographystyle{apsrev4-2} % [REVIEW] 'unsrt' is not an APS style; apsrev4-2 ships with revtex4-2.
\bibliography{ref}

\appendix

\section{Alternative decompositions with the latent instrument.}

\begin{figure*}[t!]
    \centering
    \includegraphics[width=\textwidth]{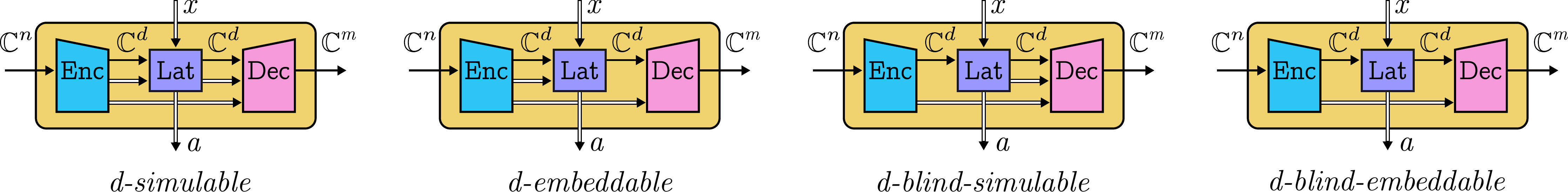}
    \caption{The four decompositions of a set of instruments $\{\mathcal{I}_{a|x}\}$ featuring a latent instrument acting on the $d$-dimensional system. They differ only in which classical variables are exchanged between the encoder, the latent instrument and the decoder. Concretely, the ``blind'' decompositions are obtained by forbidding the auxiliary classical communication between the encoder and the latent instrument.
    }
    \label{fig: variants} 
\end{figure*}

 \label{app: d-imbed}
The $d$-simulability and $d$-embeddability definitions of the bottleneck-dimension were obtained by looking for decompositions featuring a latent instrument that operates on a quantum system of dimension $d$. They only differ in  the classical information that is allowed to be exchanged between the latent instrument, the encoder and the decoder. An attentive reader has probably noticed that in the same spirit one can define two further variants, illustrated in Fig.~\ref{fig: variants}.  Formally, these lead to the definitions of a \textit{\textbf{$\bm d$-blind-embeddable}} 
\begin{align}
\label{eq: set blindemb} 
\mathcal{I}_{a|x} &=  \sum_\lambda {\tt Dec}_{|\lambda} \circ {\tt Lat}_{a|x} \circ {\tt Enc}_{\lambda}.
\end{align}
and 
\textit{\textbf{$\bm d$-blind-simulable}} 
\begin{align}
\label{eq: set blindsim}
\mathcal{I}_{a|x} &=  \sum_\lambda {\tt Dec}_{|\lambda} \circ {\tt Lat}_{a,\lambda|x} \circ {\tt Enc}.
\end{align}
sets of instruments $\{\mathcal{I}_{a|x}\}$. In both cases the classical communication from the encoder to the latent instrument is forbidden. Hence the latter is agnostic of the $d$-dimensional subspace on which it acts, which seems to make these definitions less appealing. Nevertheless let us quickly show how they compare to the notions discussed in the main text.

First, note that both definitions are obtained from $d$-simulability by limiting the allowed classical communication. Hence they imply $d$-simulability, i.e.~Eqs.~(\ref{eq: set blindemb},\ref{eq: set blindsim})$\implies$ Eq.~\eqref{eq: set simmulable 2}. For the same reason $d$-blind-embeddability implies $d$-blind-simulability and $d$-embeddability, i.e.~Eq.~\eqref{eq: set blindemb}$\implies$ Eqs.~(\ref{eq: set embedable},\ref{eq: set blindsim}). 
Conversely, to show that all the four notions of Fig.~\ref{fig: variants} are non-equivalent, we now construct the appropriate examples for the case $d=1$.

First, consider the sets of quantum states, i.e.~instruments with trivial input dimension $n=1$. Since they do not have a quantum input, any set of states is trivially $1$-simulable and $1$-blind-simulable. On the other hand, it is not difficult to see that any set of two (different) quantum states is neither $1$-embeddable nor $1$-blind-embeddable, since these decompositions forbid any causal relation between the input label $x$ and the output quantum state. This shows the strict inclusions $1$-simulable, $1$-blind-simulable $\supsetneq$ $1$-embeddable, $1$-blind-embeddable.  

Second, consider the singleton set consisting of a single quantum measurement, i.e.~an instrument with a trivial output dimension $m=1$. It is trivially $1$-simulable and $1$-embeddable, since the measurement can be performed by the encoder, which feeds forward the output $a$ via the classical variable. In contrast, it is neither $1$-blind-simulable nor $1$-blind-embeddable, since these decompositions do not have a causal link between the quantum input and the classical output $a$.  This shows the strict inclusions $1$-simulable, $1$-embeddable $\supsetneq$ $1$-blind-simulable, $1$-blind-embeddable.

%%%%%%%%%%%%%%%%%%%%%%%%%%%%%%%%%%%%%%%%%%%%%%%
%%%%%%%%%%%%%%%%%%%%%%%%%%%%%%%%%%%%%%%%%%%%%%%
\section{The set of all noisy unitary channels}

\subsection{Embeddability for invariant instruments}
\label{sec:proof invariance}

Consider a set of instruments $\mathds{I}=\{\{\mathcal{I}_{a|x}\}_a\}_x$, with $\mathcal{I}_{a|x}: L(\mathds{C}^n)\to L(\mathds{C}^m)$. We say that the set is {\it input-invariant} if for all $\bar x$ and any unitary channel $\mathcal{V}_i$ on $L(\mathds{C}^n)$ it holds that 
\begin{equation}
    \{\mathcal{I}_{a|\bar x}\circ \mathcal{V}_i\}_a \in \mathds{I},
\end{equation}
here and below instruments are identified up to permutations of outputs. Similarly, we say that set is {\it output-invariant} if for all $\bar x$ and any unitary channel $\mathcal{V}_o$ on $L(\mathds{C}^m)$ it holds that 
\begin{equation}
    \{\mathcal{V}_o \circ \mathcal{I}_{a|\bar x}\}_a \in \mathds{I}.
\end{equation}
Finally, we say that $\mathds{I}$ is {\it input-output invariant } if
\begin{equation}
    \{\mathcal{V}_o \circ \mathcal{I}_{a|\bar x}\circ \mathcal{V}_i\}_a \in \mathds{I},
\end{equation}
for any $\bar x$,$\mathcal{V}_o $ and $\mathcal{V}_i$.

We now assume that set $\mathds{I}$ is $d$-embeddable
\be
\mathcal{I}_{a|x} =  \sum_\lambda {\tt Dec}_{|\lambda} \circ {\tt Lat}_{a|\lambda,x} \circ {\tt Enc}_{\lambda},
\ee
and show that its invariance imply a specific structure for the encoder and the decoder operations. First, assume that the set is input-invariant. Meaning that for each $\{\mathcal{I}_{a|x}\}_a$ and each $\mathcal{V}_i[\cdot]= V_i\cdot V_i^\dag$ the instrument 
\begin{equation}
    \{\mathcal{I}_{a|x'}\}_a := \{\mathcal{I}_{a|x}\circ \mathcal{V}_i\}_a
\end{equation}
with $x'=x'(x,V_i)$ is also in the set, and satisfies 
\be
\mathcal{I}_{a|x'} =  \sum_\lambda {\tt Dec}_{|\lambda} \circ {\tt Lat}_{a|\lambda,x'} \circ {\tt Enc}_{\lambda}.
\ee
Then, consider an alternative decomposition with  following (continuous-valued) instruments
\begin{align}
\overline{\tt Enc}_{\lambda, V_i}&:=  {\tt Enc}_\lambda \circ \mathcal{V}_i^\dag \\
\overline{\tt Lat}_{a|\lambda,V_i,x} &:= {\tt Lat}_{a|\lambda,x'(x,V_i)}
\end{align}
where $V_i$ is sampled from the Haar measure $\dd V_i$ and the classical variable is the combination $(\lambda,V_i)$. We have
\begin{align}
&\int \dd V_i \sum_\lambda {\tt Dec}_{|\lambda} \circ \overline{\tt Lat}_{a|\lambda,V_i,x} \circ \overline{\tt Enc}_{\lambda, V_i}\\
&= \int \dd V_i \left(\sum_\lambda {\tt Dec}_{|\lambda} \circ \overline{\tt Lat}_{a|\lambda,V_i,x} \circ {\tt Enc}_{\lambda}\right) \circ \mathcal{V}_i^\dag \\
& = \int \dd V_i \left(\sum_\lambda {\tt Dec}_{|\lambda} \circ {\tt Lat}_{a|\lambda,x'(x,V_i)} \circ {\tt Enc}_{\lambda}\right) \circ \mathcal{V}_i^\dag \\
& = \int \dd V_i  (\mathcal{I}_{a|x'})\circ \mathcal{V}_i^\dag  
=\int \dd V_i  (\mathcal{I}_{a|x} \circ \mathcal{V}_i)\circ \mathcal{V}_i^\dag = \mathcal{I}_{a|x}.
\end{align}
To avoid notation clutter we were on purpose sloppy with the possible permutations of the outputs $a$. This shows that the new operations 
${\tt Dec}_\lambda$, $\overline{\tt Lat}$ and $\overline{\tt Enc}$ also decompose the instrument set.

Next,  assume that the set is $d$-embeddable and output-invariant. Meaning that for each $\{\mathcal{I}_{a|x}\}_a$ and each $\mathcal{V}_o[\cdot]= V_o\cdot V_o^\dag$ the instrument 
\begin{equation}
    \{\mathcal{I}_{a|x''}\}_a := \{\mathcal{V}_o\circ \mathcal{I}_{a|x}\}_a
\end{equation}
with $x''=x''(x,V_o)$ is also in the set, and satisfies 
\be
\mathcal{I}_{a|x''} =  \sum_\lambda {\tt Dec}_{|\lambda} \circ {\tt Lat}_{a|\lambda,x''} \circ {\tt Enc}_{\lambda}.
\ee
In this case we define another alternative (continuous-valued) decomposition
\begin{align}
\widetilde{\tt Enc}_{\lambda, V_o}&:=  {\tt Enc}_\lambda\\
\widetilde{\tt Lat}_{a|\lambda,V_o,x} &:= {\tt Lat}_{a|\lambda,x''(x,V_o)}\\
\widetilde{\tt Dec}_{\lambda, V_o}&:=  \mathcal{V}_o^\dag \circ {\tt Dec}_\lambda
\end{align}
% [REVIEW] removed a trailing \\ that produced an empty last row of the align.
where $V_o$ is sampled from the Haar measure $\dd V_o$ (by the encoder) and the classical variable is the combination $(\lambda,V_o)$. Repeating the above calculation we find 
\begin{align}
&\int \dd V_o \sum_\lambda \widetilde {\tt Dec}_{|\lambda,V_o} \circ \widetilde{\tt Lat}_{a|\lambda,V_o,x} \circ \widetilde{\tt Enc}_{\lambda, V_o}\\
& =\dots= \int \dd V_o  \mathcal{V}_o^\dag \circ  (\mathcal{I}_{a|x''})=
\mathcal{I}_{a|x},
\end{align}
and confirm that the new operations 
$\widetilde{\tt Dec}_\lambda$, $\widetilde{\tt Lat}$ and $\widetilde{\tt Enc}$ also decompose the instrument set.

Finally, for a $d$-embeddable input-output-invariant set $\mathds{I}$, each instument 
\begin{equation}
    \{\mathcal{I}_{a|\hat x}\}_a := \{\mathcal{V}_o\circ \mathcal{I}_{a|x}\circ \mathcal{V}_i\}_a
\end{equation}
with $\hat x=\hat x(x,V_o,V_i)$ is also in the set. We can combine both ideas and show that it is also $d$-embeddable with the operations
\begin{align}\label{eq: inv Vi}
\widehat{\tt Enc}_{\lambda, V_o,V_i}&:=  {\tt Enc}_\lambda \circ \mathcal{V}_i^\dag\\
\widehat{\tt Lat}_{a|\lambda,V_o,V_i,x} &:= {\tt Lat}_{a|\lambda,\hat x(x,V_o,V_i)}\\
\widehat{\tt Dec}_{\lambda, V_o,V_i}&:=  \mathcal{V}_o^\dag \circ {\tt Dec}_\lambda. \label{eq: inv Vo}
\end{align}

\begin{widetext}

\subsection{Proof of optimality of $\cE_U$.}
\label{sec: optimality of cEU}
We now consider the general symmetric  strategy for the set of noisy unitary channels. Without  loss of generality it can be taken of the form 
\begin{align}
\widehat{\tt Enc}_{\lambda, V_o,V_i}:=  {\tt Enc}_\lambda \circ \mathcal{V}_i, \quad 
 {\tt Lat}_{|\lambda,V_o,V_i,U}, \quad 
\widehat{\tt Dec}_{|\lambda, V_o,V_i}&:=  \mathcal{V}_o\circ {\tt Dec}_{|\lambda},
\end{align}
with the unitary channels $\mathcal{V}_i$ and $\mathcal{V}_o$ sampled from the Haar measure. In comparison with Eqs.~(\ref{eq: inv Vi},\ref{eq: inv Vo})  we here removed the daggers with a change of variables in order to avoid notation clutter. Let us compute the Choi-Jamiolkowski fidelity of the resulting channel, denoted $\tilde \cE_U$, with the ideal one. For simplicity we consider the case $U=\id$. Denote $\Phi^+$ the projector on the maximally entangled state $\ket{\Phi^+}= \frac{1}{\sqrt{n}} \sum_{k=1}^n \ket{k,k}$ we obtain
\begin{align}
    F_{\rm CJ}^2(\tilde{\cE}_{\id},{\rm id})&=
\tr \left(\frac{ J_{\rm id} }{n} \frac{J_{\tilde \cE_U}}{n}\right) = \tr (\Phi^+ ({\rm id}\otimes \tilde{\cE}_{\id})[\Phi^+])\\
&=\! \int \!\dd V_i  \dd V_o \sum_\lambda \tr \left(\Phi^+ ({\rm id} \otimes \mathcal{V}_o \circ {\tt Dec}_{|\lambda} \circ {\tt Lat}_{|\lambda,V_o,V_i,\id}  \circ {\tt Enc}_\lambda \circ \mathcal{V}_i)[\Phi^+] \right)
\end{align}
Without loss of generality we can take the Kraus representation $\{D_{j|\lambda}\}$,  $\{L_{i|\lambda,V_o,V_i,\id}\}$ and $\{E_\lambda\}$ for our operations, which allows us to write
\begin{align}
    F_{\rm CJ}^2(\tilde{\cE}_{\id},{\rm id})&=\frac{1}{n^2}\! \int \!\dd V_i  \dd V_o \sum_{i,j,\lambda} \left| \tr \left(V_o D_{j|\lambda} \Pi^{(d)}L_{i|\lambda,V_o,V_i,\id}\Pi^{(d)} E_\lambda V_i)\right)\right|^2 \\
    &=\frac{1}{n^2}\! \int \!\dd V_i  \dd V_o \sum_{i,j,\lambda} \left| \tr \left(L_{i|\lambda,V_o,V_i,\id}\Pi^{(d)} E_\lambda V_i V_o D_{j|\lambda} \Pi^{(d)})\right)\right|^2 \\
    &\leq \frac{1}{n^2}\int \!\dd V_i  \dd V_o \sum_{j,\lambda} \left\|\Pi^{(d)} E_\lambda V_i V_o D_{j|\lambda} \Pi^{(d)}\right\|_*^2 = \frac{1}{n^2}\int \!\dd V \sum_{j,\lambda} \left\|\Pi^{(d)} E_\lambda V D_{j|\lambda} \Pi^{(d)}\right\|_*^2
\end{align}
where we used $\| A\|_*^2\geq \sum_i \|L_i A\|_*^2$ for a CPTP map $\sum_i L_i \cdot L_i^\dag$. This inequality can be seen by considering the operator $ V A = \sum_i L_i A \otimes \ket{i} $ with $(VA)^\dag (VA) = A^\dag \sum_i L_i^\dag L_i A = A^\dag A$ guaranteeing $\|VA\|_*=\|A\|_*$, and using 
\begin{align}
  \|A\|_*=  \|V A \|_* = \tr \sqrt{(VA)(VA)^\dag} = \tr \sqrt{\sum_i L_i AA^\dag L_i^\dag \otimes \ketbra{i}{i}} = \sum_i \|L_i A\|_* \geq \sqrt{\sum_i \|L_i A\|_*^2}.
\end{align}

Next, we can decompose the Kraus operators as $E_\lambda = E_\lambda^{(d)} U_\lambda$ and $D_{i|\lambda} = U'_{i|\lambda} D_{i|\lambda}^{(d)}$, where $E_\lambda^{(d)}$ and $D_{i|\lambda}^{(d)}$ are positive semidefinite operators supported on the $d$-dimensional subspace $\Pi^{(d)}\mathds{C}^n$, and $U_\lambda$, $U'_{i|\lambda}$ are unitaries on $\mathds{C}^n$. In addition we must have
$\sum_i (D_{i|\lambda}^{(d)})^2 = \id$  and $\tr \sum_\lambda (E_\lambda^{(d)})^2=n$.
This allows us to rewrite
\begin{align}
    F_{\rm CJ}^2(\tilde{\cE}_{\id},{\rm id})&\leq \frac{1}{n^2}\int \!\dd V \sum_{j,\lambda} \left\|\Pi^{(d)} E_\lambda V D_{j|\lambda} \Pi^{(d)}\right\|_*^2 = \frac{1}{n^2}\int \!\dd V \sum_{j,\lambda} \left\| E_\lambda^{(d)} V D_{j|\lambda}^{(d)} \right\|_*^2 
    \leq \frac{1}{n^2}\int \!\dd V \sum_{\lambda} \left\| E_\lambda^{(d)} V   \Pi^{(d)} \right\|_*^2,
\end{align}
 using again used $\| A\|_*^2\geq \sum_j \| A D_j\|_*^2$. Finally, for the remaining expression, introducing a Haar-random matrix $U^{(d)}$ on $\mathds{C}^{d}$, we get 
 \begin{align}
 \int \!\dd V \left\| E_\lambda^{(d)} V   \Pi^{(d)} \right\|_*^2 = 
 \int \!\dd V \dd U^{(d)} \left\| E_\lambda^{(d)} U^{(d)} V   \Pi^{(d)} \right\|_*^2,
\end{align}
and notice that the Kraus operators $\{\frac{\sqrt{d}}{\sqrt{\tr (E_\lambda^{(d)2}) }} E_\lambda^{(d)} U^{(d)}\}$ define a CPTP map on $L(\mathds{C}^{d})$. This allows us to get the final inequality
\begin{align}
 F_{\rm CJ}^2(\tilde{\cE}_{\id},{\rm id})&\leq  
 \frac{1}{n^2}\sum_\lambda \int \!\dd V \dd U^{(d)} \left\| E_\lambda^{(d)} U^{(d)} V   \Pi^{(d)} \right\|_*^2 \leq \sum_\lambda \frac{\tr (E_\lambda^{(d)2}) }{ d n^2} \int \!\dd V \left\| \Pi^{(d)} V   \Pi^{(d)} \right\|_*^2 = \frac{1}{d n}\int \!\dd V \left\| \Pi^{(d)} V   \Pi^{(d)} \right\|_*^2.
\end{align}

We demonstrated that the fidelity attained by the general encoding strategy $\tilde{\cE}_U$ can never exceed the one achieved by the symmetric encoding $\cE_U$, discussed in the main text. Hence the latter is the optimal $d$-embedding decomposition for the set of channels $\{\mathcal{U}^{(p)}\}$.

%%%%%%%%%%%%%%%%%%%%%%%%%%%%%%%%%%%%%%%%%%%%%%%%%%
%%%%%%%%%%%%%%%%%%%%%%%%%%%%%%%%%%%%%%%%%%%%%%%%%%
\subsection{Proof of channel invariance}
\label{sec: CU inv}
    Consider the channel $\mathcal{C}_U:=\mathcal{U}^\dag \circ \cE_U $ 
    \begin{align}
       \mathcal{C}_U[\cdot] &=\frac{1}{d n} \! \int \!\dd V_i \dd V_o\,  (U^\dag V_o\Pi^{(d)}W_{V_i,V_o,U}  \Pi^{(d)}V_i ) \cdot  (V_i^\dag \Pi^{(d)} W_{V_i,V_o,U}^\dag \Pi^{(d)} V_o^\dag U) \\
       &\text{where}\qquad W_{V_i,V_o,U} := {\rm arg} \,(\Pi^{(d)}V_i U^\dag V_o\Pi^{(d)}).
    \end{align}
Here, the argmax is to be read in the sense of the polar decomposition $W_{V_i,V_o,U} |\Pi^{(d)}V_i U^\dag V_o\Pi^{(d)}| =\Pi^{(d)}V_i U^\dag V_o\Pi^{(d)}$, which  is uniquely determined excepts on a set of measure zero. In the above expression, let us denote $V_o':=U^\dag V_o$, and change the integration variable. Note that by definition $\dd V_o'$ remains the Haar measure for any fixed $U$. In addition, we have 
\begin{align}
W_{V_i,V_o,U} = {\rm arg} \,(\Pi^{(d)}V_i U^\dag V_o\Pi^{(d)})   = {\rm arg} \,(\Pi^{(d)}V_i V_o'\Pi^{(d)})  =  W_{V_i,V_o',\id}, 
\end{align}
showing that $\mathcal{C}_U $ is indeed independent of $U$
\begin{align}
       \mathcal{C}_U[\cdot] &=\frac{1}{d n} \! \int \!\dd V_i \dd V_o'\,  (V_o'\Pi^{(d)}W_{V_i,V_o',\id}  \Pi^{(d)}V_i ) \cdot  (V_i^\dag \Pi^{(d)} W_{V_i,V_o',\id}^\dag \Pi^{(d)} {V_o'}^\dag)  = \mathcal{C}_\id [\cdot]. 
    \end{align}

Finally, consider the ``rotated'' channel. Using $\mathcal{C}_U=\mathcal{C}_\id$ we have
\begin{align}
      \mathcal{V}^\dag\circ\mathcal{C}_U \circ \mathcal{V}[\cdot] &=\mathcal{V}^\dag\circ\mathcal{C}_\id \circ \mathcal{V}[\cdot] \nonumber\\
       &=\frac{1}{d n} \! \int \!\dd V_i \dd V_o\,  (V^\dag  V_o\Pi^{(d)}W_{V_i,V_o,\id }  \Pi^{(d)}V_i V ) \cdot  ({\rm h.c.}).
    \end{align}
Substituting $\bar V_i :=V_iV$ and $\bar V_o:= V^\dag V_o$, which are again Haar distributed, and noting that
$\Pi^{(d)}\bar V_i \bar V_o\Pi^{(d)} = \Pi^{(d)} V_i V V^\dag V_o\Pi^{(d)} =\Pi^{(d)} V_i V_o\Pi^{(d)}$, so that $W_{V_i,V_o,\id}=W_{\bar V_i,\bar V_o,\id}$, we obtain
\begin{align}
       \mathcal{V}^\dag\circ\mathcal{C}_U \circ \mathcal{V}[\cdot]  &=\frac{1}{d n} \! \int \!\dd \bar V_i \dd \bar V_o\,  (\bar V_o\Pi^{(d)}W_{\bar V_i,\bar V_o,\id }  \Pi^{(d)}\bar V_i ) \cdot  ({\rm h.c.}) = \mathcal{C}_{\id}[\cdot] = \mathcal{C}_{U}[\cdot].
    \end{align}
All channels $\mathcal{V}^\dag\circ\mathcal{C}_U \circ \mathcal{V}=\mathcal{C}_U=\mathcal{C}_\id$ are thus invariant and must be of the specific form
\be
\mathcal{C}_U = q\,  {\rm id} +(1-q)\, \mathcal{R}_{\id/n}\;=\;\mathcal{W}_q.
\ee

\end{widetext}

%%%%%%%%%%%%%%%%%%%%%%%%%%%%%%%%%%%%%%%%%%%%%%%
%%%%%%%%%%%%%%%%%%%%%%%%%%%%%%%%%%%%%%%%%%%%%%%
\section{The embeddability threshold for $d=n-1$}
\label{app: f2 top}

The white-noise threshold~\eqref{eq: emb Up} for $d$-embeddability of the set of noisy unitary channels is expressed through
\be\label{eq: f2 def app}
\mathfrak{f}_2(d,n)= \frac{1}{d^2}\,\mathds{E}_{V\sim{\rm CUE}(n)}\Big[\big\|\Pi^{(d)} V \Pi^{(d)}\big\|_*^2\Big],
\ee
which we evaluated numerically for general $(d,n)$. In the extremal case $d=n-1$, however, an elementary closed form is easy to derive, which is what we do now.

Choose the projector $\Pi^{(n-1)}=\id-\ketbra{n}{n}$ and regard $A:=\Pi^{(n-1)} V \Pi^{(n-1)}$ as an $(n-1)\times(n-1)$ matrix. Then
\be\label{eq: rank one}
A^\dag A = \Pi^{(n-1)} V^\dag \big(\id-\ketbra{n}{n}\big) V \Pi^{(n-1)}
= \Pi^{(n-1)}-\ketbra{v}{v},
\ee
where $\ket{v}:=\Pi^{(n-1)}V^\dag\ket{n}$ is the restriction of a Haar-random vector $V^\dag\ket{n}$ on a $(n-1)$-dimensional subspace, and $\Pi^{(n-1)}$ is the identity matrix on this subspace.
%has components $v_j=\overline{V_{nj}}$, $j=1,\dots,n-1$. 
Since the $n$-th row of $V$ is normalized,
\be
\|v\|^2=\braket{v}{v}=\sum_{j=1}^{n-1}|V_{nj}|^2 = 1-|V_{nn}|^2 .
\ee
Equation~\eqref{eq: rank one} is a rank-one perturbation of the identity: $A^\dag A$ has the eigenvalue $1$ on the orthogonal complement of $\ket{v}$ inside the subspace, i.e.~with multiplicity $n-2$, and the eigenvalue $1-\|v\|^2=|V_{nn}|^2$ along $\ket{v}$. The singular values of $A$ are therefore $1$ with multiplicity $n-2$, together with $|V_{nn}|$, so that
\be\label{eq: norm top}
\|A\|_*=(n-2)+|V_{nn}| .
\ee

A single scalar average of $|V_{nn}|$ is thus all that is needed. The last column of a Haar-random unitary is a uniformly distributed unit vector of $\mathds{C}^n$, hence $t:=|V_{nn}|^2$ follows the Beta distribution~\cite{zyczkowski2000truncations} with density $(n-1)(1-t)^{n-2}$ on $[0,1]$. Consequently $\mathds{E}[t]=\frac1n$ and
\begin{align}\label{eq: gn}
g_n:=\mathds{E}\big[\sqrt{t}\big]&=(n-1)\,\mathrm{B}\big(\tfrac32,n-1\big)\nonumber\\
&=\frac{\sqrt{\pi}\,\Gamma(n)}{2\,\Gamma\!\left(n+\frac12\right)}=\frac{4^{n}\,(n-1)!\,n!}{2\,(2n)!} ,
\end{align}
which behaves as $g_n=\frac12\sqrt{\pi/n}\,\big(1+O(n^{-1})\big)$. Squaring Eq.~\eqref{eq: norm top} and averaging we obtain
\be\label{eq: f2 top}
\mathfrak{f}_2(n-1,n)=\frac{(n-2)^2+2(n-2)\,g_n+\frac1n}{(n-1)^2} ,
\ee
or, equivalently, $\mathds{E}[\lambda_1\lambda_2]=\big[(n-3)+2g_n\big]/(n-1)$ for the cross-moment appearing in the main text. It is a rational number, e.g.~$\mathfrak{f}_2(2,3)=3/5$, $\mathfrak{f}_2(3,4)=851/1260$ and $\mathfrak{f}_2(7,8)=685441/840840$, reproducing the last diagonal of Table~\ref{tab: tab1}.

Inserting Eq.~\eqref{eq: f2 top} into Eq.~\eqref{eq: emb Up} yields the exact white-noise threshold for th case $d=n-1$
\begin{align}\label{eq: pE top}
1-p_{\rm E}^*(n-1,n)&=\frac{n\left(\,3n-4-2(n-2)g_n-\frac1n\,\right)}{(n-1)\,(n^2-1)}\\
&= \frac{3}{n}-\frac{\sqrt{\pi}}{n^{3/2}}+O(n^{-2}) .
\end{align}
% to be compared with $1-p_{\rm S}^*(n-1,n)=\frac{n}{n^2-1}=\frac1n+O(n^{-3})$. The set of noisy unitary channels thus becomes $(n-1)$-simulable already at $p\simeq 1-\frac1n$, but only $(n-1)$-embeddable at $p\simeq 1-\frac3n$. The convergence in Eq.~\eqref{eq: pE top asym} is slow --- $n\,(1-p_{\rm E}^*)\approx2.33$ at $n=8$ and $2.84$ at $n=128$ --- which accounts for the apparent $\approx 2.3/n$ scaling of the last diagonal of Table~\ref{tab: tab1}.

% Finally, the same argument applied to $d=1$ gives $A=V_{11}$ and $\mathfrak{f}_2(1,n)=\mathds{E}\big[|V_{11}|^2\big]=\frac1n$, hence $p_{\rm E}^*(1,n)=0$: a one-dimensional bottleneck can only reproduce the set $\{\mathcal{U}^{(p)}\}$ when the channels are completely depolarizing, in agreement with the first column of Table~\ref{tab: tab1}.

\end{document}